\documentclass[twocolumn, twocolappendix]{aastex631}
\usepackage{amsmath}
\usepackage{float}
\usepackage{graphicx}
\usepackage[caption=false]{subfig}
\usepackage{import}

\begin{document}
\title{Mitigating the binary viewing angle bias for standard sirens}

\correspondingauthor{Alberto Salvarese}
\email{alberto.salvarese@utexas.edu}

\author[0000-0001-9653-1778]{Alberto Salvarese} 
\author[0000-0001-5403-3762]{Hsin-Yu Chen} 
\affiliation{Department of Physics, The University of Texas at Austin, 2515 Speedway, Austin, TX 78712, USA}

\begin{abstract}
The inconsistency between experiments in the measurements of the local Universe expansion rate, the Hubble constant, suggests unknown systematics in the existing experiments or new physics. Gravitational-wave standard sirens, a method to independently provide direct measurements of the Hubble constant, have the potential to address this tension. Before that, it is critical to ensure there is no substantial systematics in the standard siren method. A significant systematic has been identified when the viewing angle of the gravitational-wave sources, the compact binary coalescences, is inferred inaccurately from electromagnetic observations of the sources. Such systematic has led to more than 10\% discrepancy in the standard siren Hubble constant measurements with the observations of binary neutron star merger, GW170817. In this Letter, we develop a new formalism to infer and mitigate this systematic. We demonstrate that the systematic uncertainty of the Hubble constant measurements can be reduced to smaller than their statistical uncertainty with 5, 10, and 20 binary neutron star merger observations. We show that our formalism successfully reduces the systematics even if the shape of the biased viewing angle distribution does not follow precisely the model we choose. Our formalism ensures unbiased standard siren Hubble constant measurements when the binary viewing angles are inferred from electromagnetic observations.   
\end{abstract}


\section{Introduction}

The expansion rate of the local Universe, the Hubble constant ($H_0$), is one of the most important constants in modern cosmology. Despite numerous observational efforts, a statistically significant discrepancy persists between direct measurements and indirect inferences of $H_0$ \citep{kamionkowski2022hubble, Planck:2018vyg, BOSS:2016wmc, Riess:2016jrr, Riess_2022, Ross:2014qpa, DiValentino:2021izs, Verde:2019ivm}. For example, the tension between the Hubble constant measured by SH0ES team using Type Ia supernovae [$H_0 = 73.3 \pm 1.04$ km/s/Mpc; \citet{Riess_2022}] and inferred by Planck collaboration using cosmic microwave background [$H_0 = 67.4 \pm 0.5$ km/s/Mpc; \citet{Planck:2018vyg}] is as large as $\sim 8\%$. 

Gravitational wave (GW) observations of compact binary coalescences (CBCs) offer an independent and direct measurement of $H_0$. The luminosity distance to CBCs can be inferred from the amplitude of the GW signals. When combining with the redshift estimates, we can measure the cosmological parameters \citep{1986Natur.323..310S}. This is known as the `standard siren' method. 
Nearby CBCs ($z\lesssim 0.1$) are ideal for the measurement of $H_0$.
There are several possibilities to estimate the redshift of CBCs. 
If an electromagnetic (EM) counterpart of a CBC is observed, it is possible to precisely localize the CBC's host galaxy, and determine the redshift from spectral follow-up or existing galaxy catalogs \citep{1986Natur.323..310S, Holz_2005}. 
With about 50 CBCs and their EM counterparts, $H_0$ can be determined to $\sim 2\%$ precision, shedding light on the Hubble tension \citep{1986Natur.323..310S, Chen_2018,nissanke2013determining,Feeney_2019}. 

\begin{figure}[h!]
    \centering
    \includegraphics[width=0.45\textwidth]{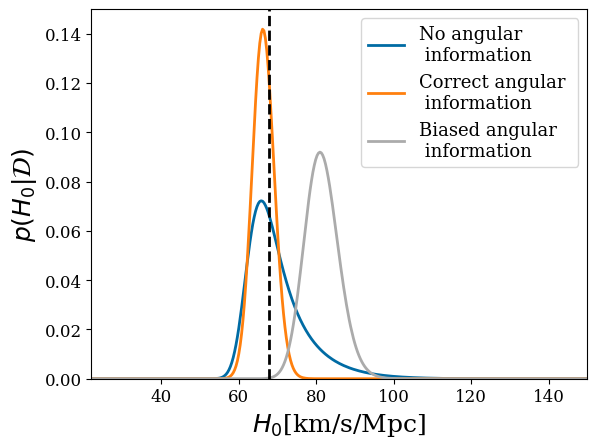}
    \caption{An example of standard siren $H_0$ posterior by combining three simulated 1.4-1.4$M_{\odot}$ binary neutron star mergers detected by LIGO-Virgo. The blue curve assumes no extra information about the binary viewing angle. The orange curve assumes the viewing angles are constrained accurately with $1\sigma$ uncertainty of $5^{\circ}$, and the gray curve assumes the viewing angles are overestimated by $20^{\circ}$ (with $1\sigma$ statistical uncertainty of $5^{\circ}$). The vertical dashed line indicates the $H_0$ value chosen for the simulations.}
    \label{fig: systematiceffect}
\end{figure}

Furthermore, it is known that the precision of standard siren measurements can be improved by constraining CBC's viewing angle ($\iota$) \citep{ChenViewingAngle, Margutti2017, Wu_2018, Mooley_2018, evans, GW170817_GWEM_angle}. GW emissions from a CBC are anisotropic. The emissions are stronger along the rotational axis of binaries.
Therefore, signals received from a faraway face-on binary are similar to those from a nearby edge-on binary \citep{Abbott_2019}. By constraining the viewing angle, the degeneracy between binary distance and viewing angle can be broken, tightening the estimate of luminosity distance, and reducing the uncertainty of $H_0$ measurements \citep{ChenViewingAngle}. Nicely, the observations of EM counterparts not only provide the redshift estimate but also allow for the measurement of CBC's viewing angle. The observations of the jet and afterglow of gamma-ray burst and kilonova following the binary neutron star (BNS) merger, GW170817 \citep{gw170817, Abbott_2017}, have been used to constrain the binary's viewing angle ~\citep{Mooley_2018, evans, GW170817_GWEM_angle,peng2024kilonova} and improve the standard siren $H_0$ measurements~\citep{Guidorzi_2017, hotokezaka2018hubble, Dhawan_2020, palmese2023standard}. In Fig. \ref{fig: systematiceffect}, we give an example of the standard siren $H_0$ measurement by combining three simulated BNS observations with (orange) and without (blue) viewing angle constraints. The $H_0$ measurement uncertainty with viewing angle constraints is substantially smaller. 

However, these constraints on viewing angle are EM model-dependent, and the methods and results of different existing analyses remain to be cross-checked. Biased viewing angle constraints can lead to significant bias in $H_0$~\citep{Chen_2020}. For example, the estimate of the viewing angle of GW170817 varied from $22^{\circ}$ to $50^{\circ}$~\citep{Guidorzi_2017, Mooley_2018, hotokezaka2018hubble, Dhawan_2020, Wang_2021, GW170817_GWEM_angle, 10.1093/mnras/stab221,palmese2023standard}, leading to $H_0$ measurements differed by more than 10\% (e.g., $H_0 = 68.3^{+4.6}_{-4.5}$ km/s/Mpc by~\cite{Mooley_2018} and $H_0 = 75.5^{+11.6}_{-9.6}$ km/s/Mpc by~\cite{Guidorzi_2017}). This is a discrepancy larger than the difference between other $H_0$ measurements, making it impossible to resolve the tension. 
In Fig. \ref{fig: systematiceffect}, we give an example of the $H_0$ measurement assuming $20^{\circ}$ overestimate of the viewing angle (gray), showing the impact of the bias. The impact is expected to become more significant as the number of observations increases.

In this Letter, we develop a new method to mitigate this bias by using the viewing angle estimated from GW signals. Although the viewing angles are often poorly constrained in GWs~\citep{ChenViewingAngle}, the measurements are well-established and are expected to be unbiased \citep{Abbott_2019}. Therefore, we can use the GW-inferred viewing angles from multiple observations to reveal the bias in EM observations. In the following, we first lay out our formalism. We then simulate BNSs detected by LIGO and Virgo, and assume the binary viewing angles are measured inaccurately from EM observations with four different types of bias. We use these simulations to show how the biases affect the standard siren $H_0$ measurements. Finally, we demonstrate that our formalism successfully reduces the systematics to less than the statistical uncertainty of the $H_0$ measurements.

\section{Mitigating the binary viewing angle bias}
Although GW detectors are able to measure the angle between the total angular momentum of the binary ($\Vec{J}$) and the line of sight ($\Vec{N}$), the so-called \textit{inclination angle}, $\theta_{\rm JN}$, most EM observations only infer the binary \textit{viewing angle}, $\iota$, which is defined as $\iota \equiv {\rm min}(\theta_{\rm JN},180^{\circ}-\theta_{\rm JN})$.

For a binary with viewing angle $\iota_*$, we assume the EM data suggests the viewing angle to be $\iota_*+b$, where $b$ is the amount of bias. 
The bias is not necessarily the same for different BNS events. Suppose the biases among different events follow an unknown underlying probability density distribution, we parametrize this distribution with a vector of parameters, $\vec{\beta}$.
Following this parameterization, we jointly infer $H_0$ and $\vec{\beta}$ using Bayesian inference, and write the posterior of $(H_0,\vec{\beta})$ as
\begin{equation}\label{eq:bayes}
    p(H_0,\vec{\beta}|\vec{\mathcal{D}}) = \pi(H_0,\vec{\beta})\frac{\mathcal{L}(\vec{\mathcal{D}}|H_0,\vec{\beta})}{p(\vec{\mathcal{D}})},
\end{equation}
where $\pi(H_0,\vec{\beta})$ stands for the prior on $(H_0,\vec{\beta})$. $\mathcal{L}(\vec{\mathcal{D}}|H_0,\vec{\beta})$ is the likelihood function, and $p(\vec{\mathcal{D}})$ is the evidence. $\Vec{\mathcal{D}}\equiv({\vec{\mathcal{D}}_{\rm GW}},\vec{\mathcal{D}}_{\rm EM})$ denotes the data from GW and EM observations. When there are $N$ BNS events, $\Vec{\mathcal{D}}$ represents the collection of data $\Vec{\mathcal{D}}=(\vec{\mathcal{D}}_{\rm GW}^1,\vec{\mathcal{D}}^1_{\rm EM},\vec{\mathcal{D}}_{\rm GW}^2,\vec{\mathcal{D}}^2_{\rm EM},...,\vec{\mathcal{D}}_{\rm GW}^N,\vec{\mathcal{D}}^N_{\rm EM})$. Assuming each BNS observation is independent, we can rewrite Eq.~\ref{eq:bayes} as
\begin{equation}
    p(H_0,\vec{\beta}|\vec{\mathcal{D}}) \propto \pi(H_0,\vec{\beta}) \prod_{i=1}^N \mathcal{L}(\vec{\mathcal{D}}_{\rm GW}^i,\vec{\mathcal{D}}^i_{\rm EM}|H_0, \vec{\beta}).
    \label{eq: posterior}
\end{equation}

If we use $\vec\Theta$ to denote the collection of binary physical parameters, such as luminosity distance $D_L$, redshift $z$, inclination angle $\theta_{\rm JN}$, mass, and spin, we can write the likelihood of $i$th event as 
\begin{equation}
\begin{split}
&\mathcal{L}(\vec{\mathcal{D}}_{\rm GW}^i,\vec{\mathcal{D}}^i_{\rm EM}|H_0, \vec{\beta})
 = \\ &\frac{\int \mathcal{L}(\vec{\mathcal{D}}_{\rm GW}^i,\vec{\mathcal{D}}^i_{\rm EM},\vec{\Theta}|H_0, \vec{\beta})d\vec{\Theta}}{\int\displaylimits_{\substack{%
    \Vec{\mathcal{D}}_{\rm GW} > \Vec{\mathcal{D}}_{\rm GW, th}\\ \Vec{\mathcal{D}}_{\rm EM} > \Vec{\mathcal{D}}_{\rm EM, th}}} \mathcal{L}(\vec{\mathcal{D}}_{\rm GW},\vec{\mathcal{D}}_{\rm EM},\vec{\Theta}|H_0, \vec{\beta})d\vec{\Theta}d\Vec{\mathcal{D}}_{\rm GW}d\Vec{\mathcal{D}}_{\rm EM}},
    \label{eq: likelihood_main}
\end{split}
\end{equation}
where $\vec{\mathcal{D}}_{\rm GW,th}$ and $\vec{\mathcal{D}}_{\rm EM,th}$ denote the detection threshold of GW detectors and EM telescopes. The denominator of Eq.~\ref{eq: likelihood_main} accounts for the selection effect, 
since not all sources are equally detectable~\citep{Loredo_selection, Thrane_2019, mandel_2019, Vitale2020}. 

We can further separate $\vec\Theta$ into relevant physical parameters ($D_L$,$z$,$\Theta_{\rm JN}$) and other physical parameters $\vec{\Theta}^{'}$ ( $\vec{\Theta}\supset \{D_L,z,\Theta_{\rm JN},\vec{\Theta}^{'}\}$), and write the likelihood for individual event as
\begin{equation}
    \begin{split}
        \mathcal{L}(\vec{\mathcal{D}}_{\rm GW}^i, \Vec{\mathcal{D}}_{\rm EM}^i|&H_0,\vec{\beta}) \propto \int d\theta_{\rm JN} dD_{\rm L} dz d\vec{\Theta}^{'} db \\ &\times p(\vec{\mathcal{D}}_{\rm GW}^i, \Vec{\mathcal{D}}_{\rm EM}^i,\theta_{\rm JN},D_{\rm L},z,\vec{\Theta}^{'},b|H_0, \vec{\beta}).
        \label{eq: marginalized_likelihood_main}
    \end{split}
\end{equation}
Note that in Eq.~\ref{eq: marginalized_likelihood_main} we explicitly write the bias $b$ as one of the parameters that affect the data. 

Since the GW and EM observations are independent, we 
can write the integrand of Eq.~\ref{eq: marginalized_likelihood_main} as (see Appendix for step-by-step derivations)
\begin{equation}
\begin{split}
    & p(\Vec{\mathcal{D}}_{\rm GW}^i, \Vec{\mathcal{D}}_{\rm EM}^i, \theta_{\rm JN}, D_{\rm L}, z, \vec{\Theta}^{'},b|H_0,\Vec{\beta}) \\&=  
    \: \mathcal{L}(\Vec{\mathcal{D}}_{\rm GW}^i|\theta_{\rm JN}, D_{\rm L}, \vec{\Theta}^{'})
    \times \mathcal{L}(\Vec{\mathcal{D}}_{\rm EM}^i| \theta_{\rm JN}, z, \vec{\Theta}^{'},b)\\
    &\times p(\theta_{\rm JN}, D_{\rm L}, z, \vec{\Theta}^{'},b|H_0, \Vec{\beta}).
    \label{eq: separated_full_likelihood}
\end{split}
\end{equation}
Note that the GW and EM data provide the measurements of luminosity distance $D_L$ and redshift $z$, respectively, and only the EM data are affected by the viewing angle bias $b$. 

Finally, the last term in Eq.~\ref{eq: separated_full_likelihood} can be written as 
\begin{equation}
\begin{split}
    p(\theta_{\rm JN}, D_{\rm L}, z,\vec{\Theta}^{'}, b|H_0, \Vec{\beta}) =& \delta[D_{\rm L} - \hat{D}_{\rm L}(z, H_0)]\\
    & \times p(\theta_{\rm JN})p(z|H_0)p(\vec{\Theta}^{'})p(b|\Vec{\beta}),
    \label{eq: joint_quantities_pdf}
\end{split}
\end{equation}
$p(b|\Vec{\beta})$ represents the probability distribution of the viewing angle bias under a given model. In this Letter, we start with a Normal distribution as our baseline model for $p(b|\Vec{\beta})$. We will show later that this baseline model works well to mitigate the bias even if the bias \textit{does not} follow a Normal distribution. Under this model, the probability distribution is described by two parameters, $\vec{\beta}=(\beta_1,\beta_2)$, where $\beta_1$ and $\beta_2$ represent the mean and standard deviation of the Normal distribution, respectively.

$\delta[D_{\rm L} - \hat{D}_{\rm L}(z, H_0)]$ is a Dirac $\delta$-function originating from the dependency between luminosity distance, redshift, and $H_0$ when other cosmological parameters are fixed, an assumption that is valid for nearby events but can be relaxed when events at higher redshifts are included. $p(\theta_{\rm JN})$ and $p(z|H_0)$ are chosen so that the binaries are assumed to be uniformly distributed in comoving volume and have random inclination angles. 

\begin{figure}[t]
    \centering
    \includegraphics[width=0.45\textwidth]{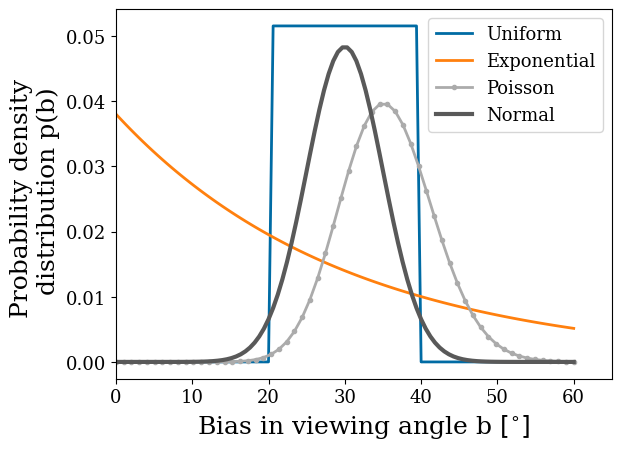}
    \caption{Probability density distribution of the bias in viewing angle we study in this Letter. The bias for each simulated BNS event is randomly drawn from these distributions.}
    \label{fig: distributions}
\end{figure} 
\begin{figure*}[ht!]
    \centering
    \includegraphics[width=0.9\textwidth]{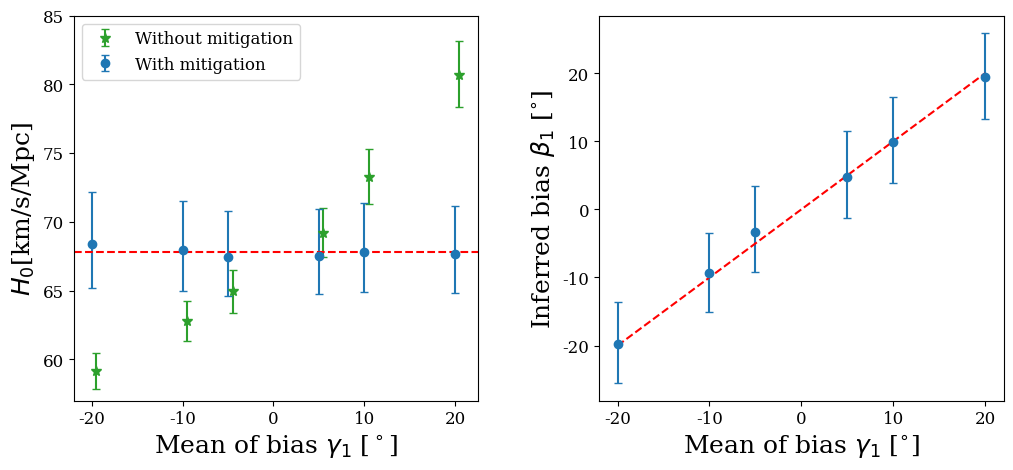}
    \caption{Measurements from 10 simulated BNS detections, averaged over 30 realizations. The horizontal axes label the mean ($\gamma_1$) of the viewing angle bias distribution. Here we assume the bias follows a Normal distribution centering at $\gamma_1$ with a standard deviation of $5^{\circ}$. Left: The median and symmetric 68\% credible interval of the $H_0$ posteriors with (blue) and without (green) applying our mitigation formalism. The red dashed line marks the simulated value of $H_0$. Right: The median and symmetric 68\% credible interval of the inferred bias ($\beta_1$) posteriors. The red dashed line represents the line of accurate measurement.}
    \label{fig: comparisons}
\end{figure*}
\section{Application to simulated observations}
We simulate 1000 1.4-1.4 M$_{\odot}$ BNS detections detected by LIGO-Hanford, LIGO-Livingston, and Virgo with their 4th observing run sensitivities (\texttt{
aligo\_O4high.txt}, \texttt{avirgo\_O4high\_NEW.txt} in ~\cite{Abbott_2020})~\citep{LIGOScientific:2014pky, VIRGO:2014yos} in a Universe following Planck 2015 cosmology~\citep{PhysRevD.107.023531} ($\Omega_m = 0.306$, $\Omega_{\Lambda}=0.694$, $\Omega_k = 0.0$, $H_0 = 67.8$km/s/Mpc). The detections were drawn from a BNS population with isotropic orientations and are distributed uniformly in comoving volume. The GW detection threshold is set at a network signal-to-noise ratio of 12 \citep{Abbott_2020, Chen_2021}. 
We follow the method developed in~\cite{ChenViewingAngle} to construct the GW likelihood function, $\mathcal{L}(\Vec{\mathcal{D}}_{\rm GW}^i|\theta_{\rm JN}, D_{\rm L}, \vec{\Theta}^{'})$ in Eq.~\ref{eq: separated_full_likelihood}. For the EM likelihood, $\mathcal{L}(\Vec{\mathcal{D}}_{\rm EM}^i| \theta_{\rm JN}, z, \vec{\Theta}^{'},b)$, 
we assume that the redshift measurements from EM observations have negligible uncertainty, an appropriate assumption when compared to the uncertainty in GW distance measurements \citep{gw170817}. For the viewing angle measured in EM observations, we add $\epsilon_*$ to each binary viewing angle $\iota_*$, where $\epsilon_*$ is randomly drawn from a zero-mean Normal distribution with a standard deviation of $5^{\circ}$, to account for the statistical uncertainty of the measurements \citep{Margutti2017, Wu_2018, Mooley_2018, evans, GW170817_GWEM_angle}. In addition, we randomly draw a bias $b_*$ for each binary from one of the probability distributions described below. Therefore, the likelihood of the viewing angle measured from EM observations is simulated as a 1-dimensional Gaussian function with a $1\sigma$ width of $5^{\circ}$ centering at $\Tilde{\iota} = \iota_*+\epsilon_*+b_*$.
To simulate the bias in viewing angle, we explore four potential probability distributions (see examples in Fig. \ref{fig: distributions}): (i) Normal distribution with mean $\gamma_1$ and standard deviation $\gamma_2$. This is the same type of distribution as our baseline model for $p(b|\Vec{\beta})$ in Eq.~\ref{eq: joint_quantities_pdf}. We pick $\gamma_1 = [\pm 5^{\circ}, \pm 10^{\circ}, \pm 20^{\circ}$], and $\gamma_2=5^{\circ}$. (ii) Uniform distribution, with supports $[-40^{\circ},0^{\circ}]$, $[-20^{\circ},0^{\circ}]$, $[-10^{\circ},0^{\circ}]$, $[0^{\circ},10^{\circ}]$, $[0^{\circ},20^{\circ}]$, and $[0^{\circ},40^{\circ}]$. (iii) Poisson distribution:
 \begin{equation}
    p(b_*) = 
     \begin{cases}
        \frac{\lambda^{b_*} e^{-\lambda}}{b_* !} \quad\mbox{if}\quad b_* \geq0\\ 
        \frac{\lambda^{-b_*} e^{-\lambda}}{(-b_*) !} \quad\mbox{if}\quad b_* <0
     \end{cases}
 \end{equation}
with expected rate parameter $\lambda = 10^{\circ}, 20^{\circ}, 30^{\circ}$. (iv) Exponential distribution: 
\begin{equation}
    p(b_*)=
    \begin{cases}
        \frac{1}{\lambda} e^{- \frac{b_*}{\lambda}} \quad\mbox{if}\quad b_*\geq 0\\
        \frac{1}{\lambda} e^{\frac{b_*}{\lambda}} \quad\mbox{if}\quad b_*<0
    \end{cases}
\end{equation} 

with scale parameter $\lambda = 10^{\circ}, 20^{\circ}, 30^{\circ}$.  
We use uniform priors $[20, 200]\mbox{km/s/Mpc}$, $[-90^{\circ}, 90^{\circ}]$, and $[2^{\circ}, 92^{\circ}-|\beta_1|]$ for $H_0$, $\beta_1$, and $\beta_2$ in Eq.~\ref{eq:bayes} respectively. The minimum for $\beta_2$ is chosen to avoid numerical effects in the Markov Chain Monte Carlo inference, and the maximum is set by the range of viewing angle $[0^{\circ},90^{\circ}]$. 

We randomly select 5, 10, and 20 events out of the 1000 BNS detections and use \texttt{emcee} \cite{Foreman_Mackey_2013} to sample the posteriors in Eq.~\ref{eq:bayes}. We repeat this process 30 times for all types of bias, and report the average of posteriors in the next section. 

\section{Results}
We start with the bias distribution that follows a Normal distribution. 
In the left panel of Fig.~\ref{fig: comparisons}, we present the median and symmetric 68\% credible interval of the $H_0$ posteriors when combining 10 BNS detections for simulated bias centering at different $\gamma_1$. 
As shown in the figure, the $H_0$ measurements significantly deviate from the simulated value even with a small $\gamma_1$ when the bias is not mitigated (green). We then show the $H_0$ posteriors following our mitigation formalism in blue (Eq.~\ref{eq:bayes} marginalized over $\vec{\beta}$). We find that our formalism successfully reduces the bias in $H_0$ to less than their statistical uncertainties for all $\gamma_1$.  
In the right panel of Fig.~\ref{fig: comparisons}, we show the $\beta_1$ posteriors marginalized over $H_0$ and $\beta_2$ for simulated bias centering at different $\gamma_1$. We find that our formalism can reveal the simulated bias accurately. 

\begin{figure}[htp]
\centering
\subfloat{%
  \includegraphics[width=0.4\textwidth]{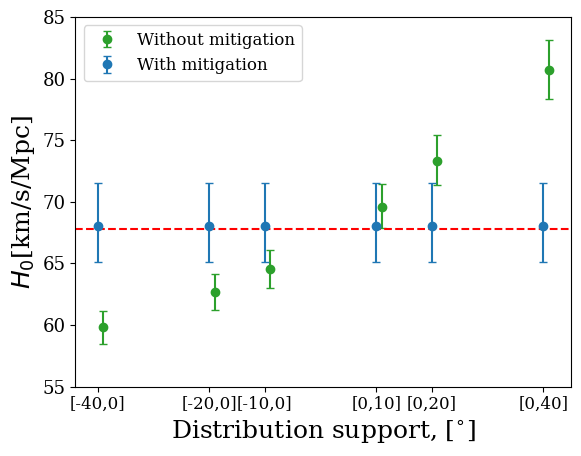}%
}

\subfloat{%
  \includegraphics[width=0.4\textwidth]{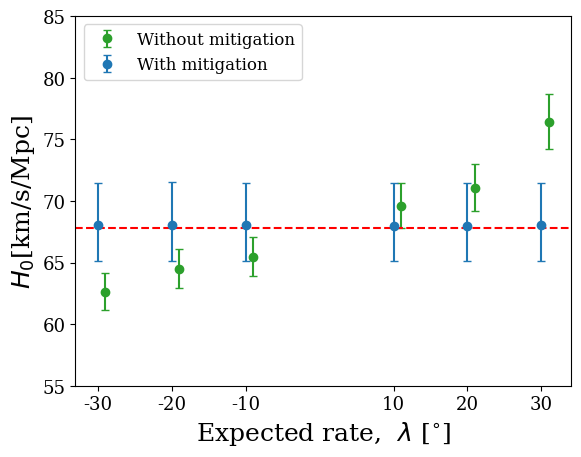}%
}

\subfloat{%
  \includegraphics[width=0.4\textwidth]{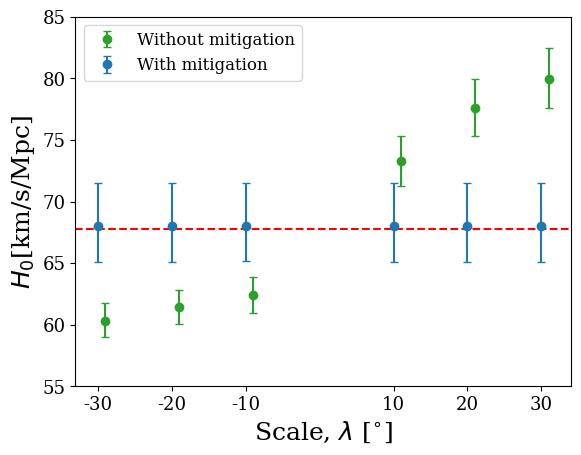}%
}
\caption{The median and symmetric 68\% credible interval of the $H_0$ posteriors with (blue) and without (green) applying our mitigation formalism. We assume the viewing angle bias follows a Uniform (top), Poisson (middle), and Exponential (bottom) distribution (see texts for the choices of distribution parameters). These distributions are \textit{different} from the parameterized model we pick for the bias. The results are measurements from 10 simulated BNS detections, averaged over 30 realizations.}
\label{fig: different_injections}
\end{figure}

In addition to $\sigma=5^{\circ}$ and $\gamma_2=5^{\circ}$, we also explore larger $\sigma$ and $\gamma_2$. Furthermore, we repeated the simulations for 5 and 20 BNS detections. We find similar results with $H_0$ and $\gamma_1$ measured accurately. 
In reality, the viewing angle bias distribution does not necessarily follow the parameterized model we pick. Therefore, we consider three additional types of bias distributions (Fig.~\ref{fig: distributions}) and explore if a Normal distribution is sufficient to mitigate the bias. In Fig. \ref{fig: different_injections}, we present the median and symmetric 68\% credible interval of the $H_0$ posteriors with (blue) and without (green) applying our mitigation formalism for these three types of bias. Even if the bias distribution does not follow a Normal distribution, we can effectively reduce the bias in $H_0$ to below the measurement statistical uncertainties when modeling the bias as a Normal distribution. This is because the central limit theorem ensures the mean of the drawn biases $b_*$ follows a Normal distribution, which can be successfully captured by our baseline model. 

We also find that the inferred mean ($\beta_1$) and standard deviation ($\beta_2$) correctly estimate the mean and standard deviation of the simulated biased distribution, even if the distribution is not Normal (please see Appendix where we present the differences).

\section{Discussion}
In this Letter, we present a new bright siren inference formalism that successfully mitigates the systematic uncertainty introduced by inaccurate binary viewing angle estimates. Our formalism ensures that the systematic uncertainty lies below the statistical uncertainty of the $H_0$ measurements when the binary viewing angle inferred from non-GW channels, such as EM observations, is biased. We show that the bias can be mitigated even if the distribution of the systematics does not precisely follow the model we assume. 

The complexity of EM emission modelings, the differences among analyses, and the variations of observing conditions could all lead to bias in the estimate of binary viewing angle. The sources of the bias can be difficult to disentangle, and the characteristics of the bias vary with the choice of data. It is therefore pivotal to demonstrate that our formalism effectively mitigates different types of bias distribution with the chosen baseline model, so that the formalism can be applied in wide-ranging circumstances.

In Figs.~\ref{fig: comparisons} and ~\ref{fig: different_injections}, we present the results for 10 simulated BNS detections. We also explore the five-detection scenario and find that five events are already informative enough to reveal and alleviate the bias with our formalism. In addition, we perform the 20-detection simulations to investigate if there is any remaining systematic after applying our formalism. The increased number of events reduces the statistical uncertainty of the measurements and the systematic uncertainty stands out. We do not find any remaining bias.

Comparing the precision of the $H_0$ measurements with (applied our mitigation formalism) and without (no need for mitigation) using the viewing angle information, we find that the precision are comparable when the number of events is low. As the number of events increases, the bias are better inferred with our formalism, and the precision of $H_0$ measurements becomes better than those without using the angular information. The transition happens around 10 and 20 events when the viewing angle bias follows a Normal and Poisson distribution, respectively. We notice that the $H_0$ precision remains to be comparable with and without using the angular information even with 40 events for Uniform and Exponential bias distributions. This is likely due to the substantial difference between the bias distributions and our baseline model (a Normal distribution). However, if
the shape of the bias distribution is better understood, more accurate models can be used to replace the Normal distribution we adopt. Even if the shape of the bias distribution is unknown, knowledge of the possible range of the bias can be used as the prior $\pi(H_0,\vec{\beta})$ in Eq.~\ref{eq:bayes} and improve the measurements. In addition, non-parametric model is another possibility to make our formalism completely model-agnostic.

\section*{Acknowledgements}
The authors would like to thank Sayantani Bera for the LIGO Scientific Collaboration internal review.
The authors are supported by the National Science Foundation under Grant PHY-2308752. The authors are grateful for computational resources provided by the LIGO Laboratory and supported by National Science Foundation Grants PHY-0757058 and PHY-0823459.
This is LIGO Document Number LIGO-P2400248.

\appendix
\section{Likelihood}
Starting from the likelihood of individual event in the main article, we apply the product rule for joint probability [$p(A,B|I) = p(A|B,I)p(B|I)$] and rewrite the integrand as
\begin{equation}
\begin{split}
    p(\Vec{\mathcal{D}}^i_{\rm GW}, \Vec{\mathcal{D}}^i_{\rm EM}&, \theta_{\rm JN}, D_{\rm L}, z, \Vec{\Theta}^{'}, b|H_0,\Vec{\beta})\\
    =\:& p(\theta_{\rm JN}, D_{\rm L}, z, \Vec{\Theta}^{'}, b|H_0, \Vec{\beta})\\
    &\times p(\Vec{\mathcal{D}}^i_{\rm GW}, \Vec{\mathcal{D}}^i_{\rm EM}|\theta_{\rm JN}, D_{\rm L}, z, \Vec{\Theta}^{'}, b, H_0, \Vec{\beta}).
    \label{eq: separated_full_likelihood_appendix}
\end{split}
\end{equation}

In the following, we repetitively apply the product rule and consider the dependencies between parameters to simplify the first line of Eq.~\ref{eq: separated_full_likelihood_appendix}:
\begin{equation}
    \begin{split}
        p(\theta_{\rm JN},& D_{\rm L}, z, \Vec{\Theta}^{'}, b|H_0, \vec{\beta})\\
        =\:&p(D_{\rm L}|\theta_{\rm JN}, z, \Vec{\Theta}^{'}, b, H_0, \vec{\beta})p(\theta_{\rm JN}, z, \Vec{\Theta}^{'}, b|H_0, \Vec{\beta})\\
        =\:& p(D_{\rm L}|z, H_0)p(\theta_{\rm JN}|z, b, H_0, \Vec{\beta}) p(z, \Vec{\Theta}^{'}, b|H_0, \Vec{\beta})\\
        =\:& p(D_{\rm L}|z, H_0)p(\theta_{\rm JN}) p(\Vec{\Theta}^{'})p(z|H_0)p(b|\Vec{\beta})\\
        =\:& \delta[D_{\rm L} - \hat{D}_{\rm L}(z, H_0)]p(\theta_{\rm JN})p(\Vec{\Theta}^{'})p(z|H_0)p(b|\Vec{\beta}).
        \label{eq: second_line}
    \end{split}
\end{equation}

The second equality originates from the fact that the intrinsic distribution of source luminosity distance is independent of the binary inclination angle $\theta_{\rm JN}$, the viewing angle bias $b$, and the bias parameters $\Vec{\beta}$. Similarly, the third equality indicates that the intrinsic distribution of binary inclination angle $\theta_{\rm JN}$ is independent of other parameters. The redshift $z$ distribution only depends on $H_0$ when other cosmological parameters are fixed. The viewing angle bias $b$ only depends on the bias distribution parameters $\vec{\beta}$. Finally, the dependency between $D_L$, $z$ and $H_0$ leads to the last line.

Since GW and EM observations are independent, we separate the second line of Eq.~\ref{eq: separated_full_likelihood_appendix} into
\begin{equation}
\begin{split}
    p(\vec{\mathcal{D}}_{\rm GW}^i, \Vec{\mathcal{D}}_{\rm EM}^i|&,\theta_{\rm JN},D_{\rm L},z, \Vec{\Theta}^{'},b,H_0,\Vec{\beta}) =  \\ &\mathcal{L}(\vec{\mathcal{D}}_{\rm GW}^i|\theta_{\rm JN},D_{\rm L},z, \Vec{\Theta}^{'},b,H_0,\Vec{\beta})\\
    &\times \mathcal{L}(\vec{\mathcal{D}}_{\rm EM}^i|\theta_{\rm JN},D_{\rm L},z, \Vec{\Theta}^{'},b,H_0,\Vec{\beta}).
\end{split}
\end{equation}

We can further simplify this expression by accounting for the dependencies of parameters. The GW data is independent of the viewing angle biases $b$ and their distribution parameters $\vec{\beta}$. Furthermore, GW data do not directly depend on $z$ or $H_0$. Therefore, 
\begin{equation}
\begin{split}
    \mathcal{L}(\Vec{\mathcal{D}}_{\rm GW}^i|&\theta_{\rm JN}, D_{\rm L},  \Vec{\Theta}^{'}, b, \Vec{\beta}) = \\
    &\mathcal{L}(\Vec{\mathcal{D}}_{\rm GW}^i|\theta_{\rm JN}, D_{\rm L}, \Vec{\Theta}^{'})
\end{split}
    \label{eq: GW_likelihood_appendix}
\end{equation}
On the other hand, EM observations provide the measurements of redshift $z$, and the inclination angle estimate is biased by the bias $b$ under our assumption:

\begin{equation}
\begin{split}
    \mathcal{L}(\Vec{\mathcal{D}}_{\rm EM}^i|&\theta_{\rm JN}, D_{\rm L}, z, \Vec{\Theta}^{'}, b) = \\
    &\mathcal{L}(\Vec{\mathcal{D}}_{\rm EM}^i|\theta_{\rm JN}, z, \Vec{\Theta}^{'}, b)
\end{split}\label{eq: EM_likelihood_appendix}
\end{equation}

We note that the EM likelihood Eq.~\ref{eq: EM_likelihood_appendix} is symmetric around $\theta_{\rm JN}=90^{\circ}$, since the EM observations only measure the \textit{viewing angle}. 

\section{Inference of the viewing angle bias}
 
\begin{figure}[ht!]
    \centering
    \includegraphics[width=0.414\textwidth]{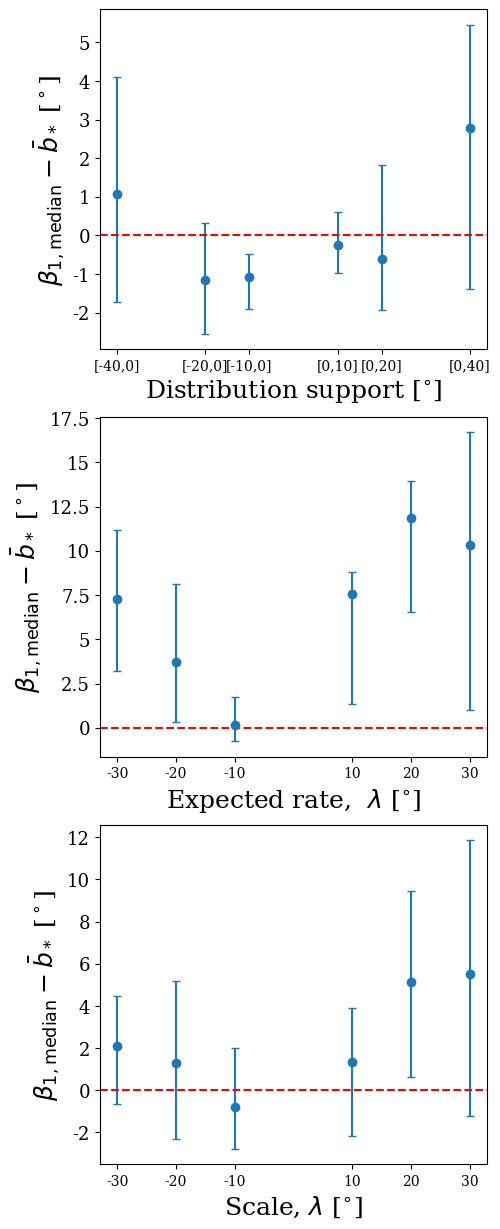}
    \caption{The difference between the median of the inferred mean (i.e., median of the $\beta_1$ posterior) and the mean of the simulated bias when combining 10 simulated BNS events. The data point and error bar show the median and symmetric 68\% credible interval of the difference for 30 realizations. The results are for the viewing angle bias following a Uniform (top), Poisson (middle), and Exponential (bottom) distribution.  }
    \label{fig: diff10}
\end{figure}

\begin{figure}[ht!]
    \centering
    \includegraphics[width=0.4\textwidth]{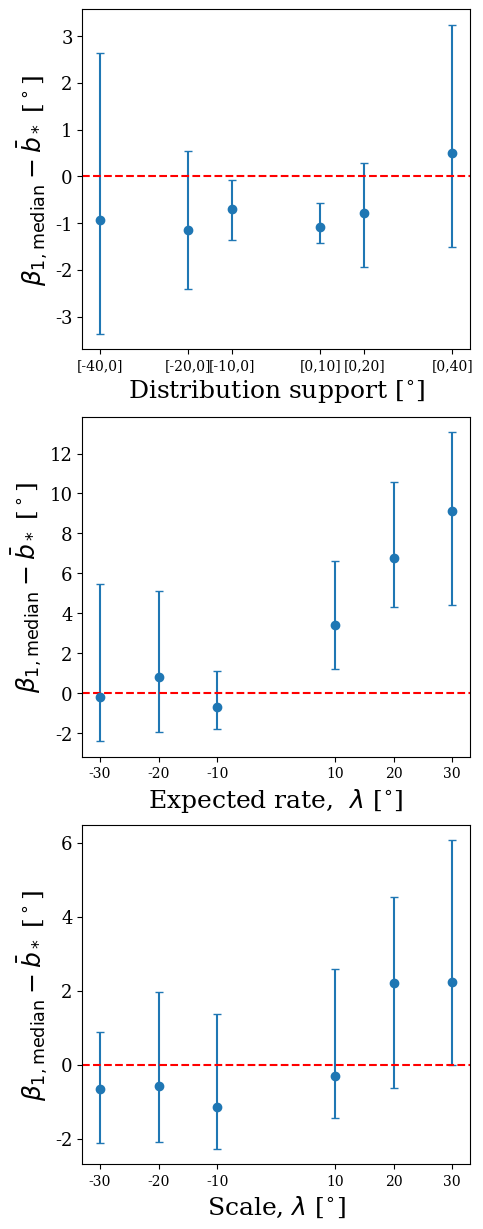}
    \caption{Same as Figure~\ref{fig: diff10}, but combining 20 simulated BNS events.}
    \label{fig: diff20}
\end{figure}

\begin{figure}[ht!]
    \centering
    \includegraphics[width=0.4\textwidth]{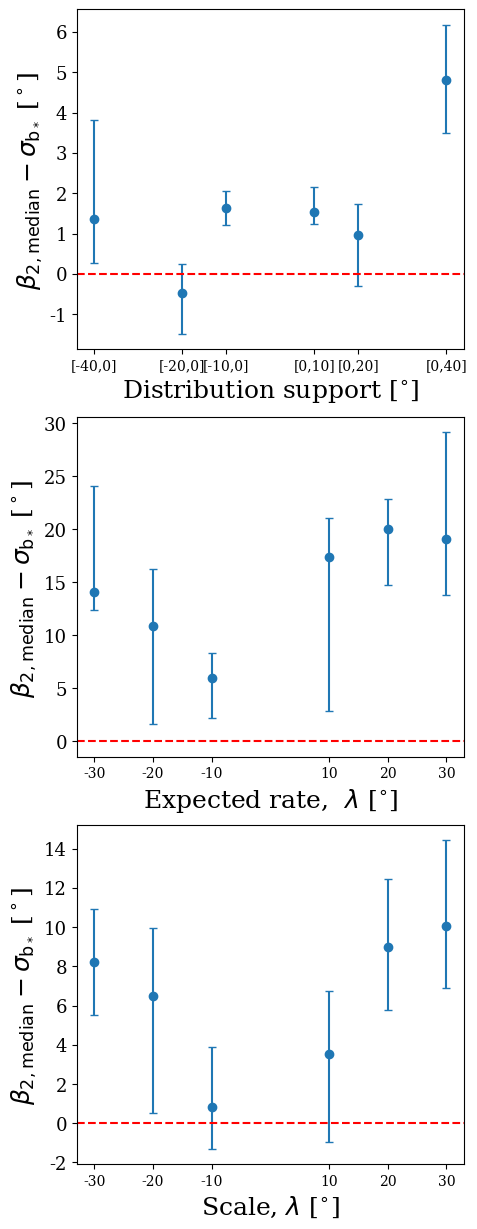}
    \caption{The difference between the median of the inferred standard deviation (i.e., median of the $\beta_2$ posterior) and the standard deviation of the simulated bias when combining 10 simulated BNS events. The data point and error bar show the median and symmetric 68\% credible interval of the difference for 30 realizations. The results are for the viewing angle bias following a Uniform (top), Poisson (middle), and Exponential (bottom) distribution.  }
    \label{fig: diff10_b2}
\end{figure}

\begin{figure}[ht!]
    \centering
    \includegraphics[width=0.41\textwidth]{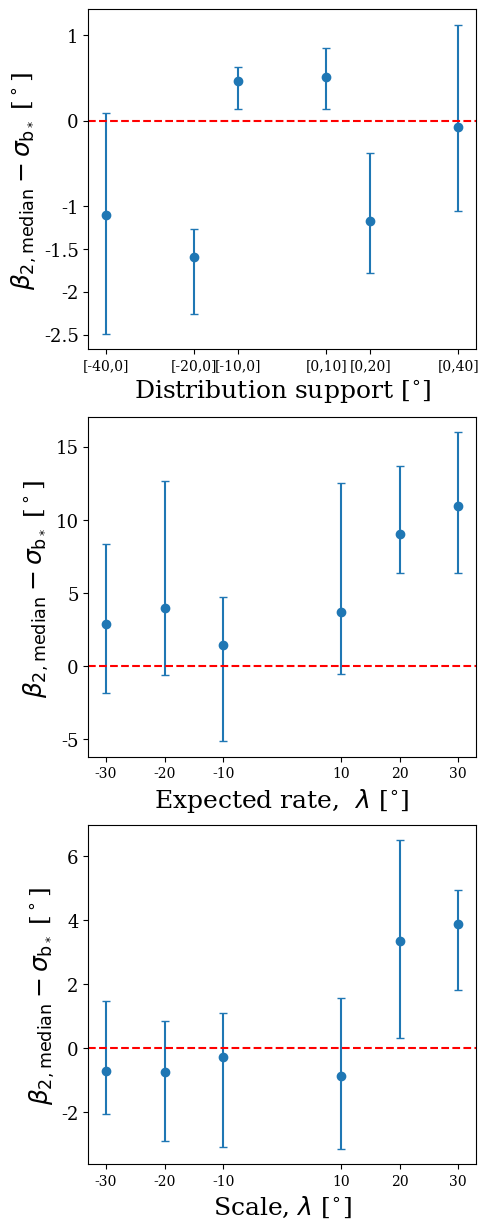}
    \caption{Same as Figure~\ref{fig: diff10_b2}, but combining 20 simulated BNS events.}
    \label{fig: diff20_b2}
\end{figure}

We compute the difference between the median of the inferred mean (i.e., median of the $\beta_1$ posterior) and the mean of the simulated bias. We repeat the computation for 30 realizations and present the median and symmetric 68\% credible interval of the difference (Figures ~\ref{fig: diff10} and ~\ref{fig: diff20}). 
Similarly, we compute the difference between the median of the inferred standard deviation (i.e., median of the $\beta_2$ posterior) and the standard deviation of the simulated bias. We then present the median and symmetric 68\% credible interval of the difference over 30 realizations (Figures ~\ref{fig: diff10_b2} and ~\ref{fig: diff20_b2}). The results are shown for all three types of non-Normally-distributed bias in the manuscript. We show the results when combining 10 and 20 events.
We find that the means of the bias distribution can be inferred fairly well with $\gtrsim 10$ events, and the standard deviations can be inferred with $\gtrsim 20$ events.

\bibliography{ref}

\begin{thebibliography}{}
\expandafter\ifx\csname natexlab\endcsname\relax\def\natexlab#1{#1}\fi
\providecommand{\url}[1]{\href{#1}{#1}}
\providecommand{\dodoi}[1]{doi:~\href{http://doi.org/#1}{\nolinkurl{#1}}}
\providecommand{\doeprint}[1]{\href{http://ascl.net/#1}{\nolinkurl{http://ascl.net/#1}}}
\providecommand{\doarXiv}[1]{\href{https://arxiv.org/abs/#1}{\nolinkurl{https://arxiv.org/abs/#1}}}

\bibitem[{Aasi {et~al.}(2015)}]{LIGOScientific:2014pky}
Aasi, J., {et~al.} 2015, Classical and Quantum Gravity, 32, 074001, \dodoi{10.1088/0264-9381/32/7/074001}

\bibitem[{Abbott {et~al.}(2017{\natexlab{a}})Abbott, Abbott, Abbott, Acernese, Adams, Adams, Addesso, Adhikari, Adya, Affeldt, Afrough, Agarwal, Agathos, Agatsuma, Aggarwal, Aguiar, Aiello, Ain, Ajith, Allen, Allen, Allocca, Altin, Amato, Ananyeva, Anderson, Anderson, Angelova, Antier, Appert, Arai, Araya, Areeda, Arnaud, Arun, Ascenzi, Ashton, Ast, Aston, Astone, Atallah, Aufmuth, Aulbert, AultONeal, Austin, Avila-Alvarez, Babak, Bacon, Bader, Bae, Bailes, Baker, Baldaccini, Ballardin, Ballmer, Banagiri, Barayoga, Barclay, Barish, Barker, Barkett, Barone, Barr, Barsotti, Barsuglia, Barta, Barthelmy, Bartlett, Bartos, Bassiri, Basti, Batch, Bawaj, Bayley, Bazzan, B\'ecsy, Beer, Bejger, Belahcene, Bell, Berger, Bergmann, Bernuzzi, Bero, Berry, Bersanetti, Bertolini, Betzwieser, Bhagwat, Bhandare, Bilenko, Billingsley, Billman, Birch, Birney, Birnholtz, Biscans, Biscoveanu, Bisht, Bitossi, Biwer, Bizouard, Blackburn, Blackman, Blair, Blair, Blair, Bloemen, Bock, Bode, Boer, Bogaert, Bohe, Bondu, Bonilla,
  Bonnand, Boom, Bork, Boschi, Bose, Bossie, Bouffanais, Bozzi, Bradaschia, Brady, Branchesi, Brau, Briant, Brillet, Brinkmann, Brisson, Brockill, Broida, Brooks, Brown, Brown, Brunett, Buchanan, Buikema, Bulik, Bulten, Buonanno, Buskulic, Buy, Byer, Cabero, Cadonati, Cagnoli, Cahillane, Calder\'on~Bustillo, Callister, Calloni, Camp, Canepa, Canizares, Cannon, Cao, Cao, Capano, Capocasa, Carbognani, Caride, Carney, Carullo, Casanueva~Diaz, Casentini, Caudill, Cavagli\`a, Cavalier, Cavalieri, Cella, Cepeda, Cerd\'a-Dur\'an, Cerretani, Cesarini, Chamberlin, Chan, Chao, Charlton, Chase, Chassande-Mottin, Chatterjee, Chatziioannou, Cheeseboro, Chen, Chen, Chen, Cheng, Chia, Chincarini, Chiummo, Chmiel, Cho, Cho, Chow, Christensen, Chu, Chua, Chua, Chung, Chung, Ciani, Ciolfi, Cirelli, Cirone, Clara, Clark, Clearwater, Cleva, Cocchieri, Coccia, Cohadon, Cohen, Colla, Collette, Cominsky, Constancio, Conti, Cooper, Corban, Corbitt, Cordero-Carri\'on, Corley, Cornish, Corsi, Cortese, Costa, Coughlin, Coughlin,
  Coulon, Countryman, Couvares, Covas, Cowan, Coward, Cowart, Coyne, Coyne, Creighton, Creighton, Cripe, Crowder, Cullen, Cumming, Cunningham, Cuoco, Dal~Canton, D\'alya, Danilishin, D'Antonio, Danzmann, Dasgupta, Da~Silva~Costa, Dattilo, Dave, Davier, Davis, Daw, Day, De, DeBra, Degallaix, De~Laurentis, Del\'eglise, Del~Pozzo, Demos, Denker, Dent, De~Pietri, Dergachev, De~Rosa, DeRosa, De~Rossi, DeSalvo, de~Varona, Devenson, Dhurandhar, D\'{\i}az, Dietrich, Di~Fiore, Di~Giovanni, Di~Girolamo, Di~Lieto, Di~Pace, Di~Palma, Di~Renzo, Doctor, Dolique, Donovan, Dooley, Doravari, Dorrington, Douglas, Dovale~\'Alvarez, Downes, Drago, Dreissigacker, Driggers, Du, Ducrot, Dudi, Dupej, Dwyer, Edo, Edwards, Effler, Eggenstein, Ehrens, Eichholz, Eikenberry, Eisenstein, Essick, Estevez, Etienne, Etzel, Evans, Evans, Factourovich, Fafone, Fair, Fairhurst, Fan, Farinon, Farr, Farr, Fauchon-Jones, Favata, Fays, Fee, Fehrmann, Feicht, Fejer, Fernandez-Galiana, Ferrante, Ferreira, Ferrini, Fidecaro, Finstad, Fiori, Fiorucci,
  Fishbach, Fisher, Fitz-Axen, Flaminio, Fletcher, Fong, Font, Forsyth, Forsyth, Fournier, Frasca, Frasconi, Frei, Freise, Frey, Frey, Fries, Fritschel, Frolov, Fulda, Fyffe, Gabbard, Gadre, Gaebel, Gair, Gammaitoni, Ganija, Gaonkar, Garcia-Quiros, Garufi, Gateley, Gaudio, Gaur, Gayathri, Gehrels, Gemme, Genin, Gennai, George, George, Gergely, Germain, Ghonge, Ghosh, Ghosh, Ghosh, Giaime, Giardina, Giazotto, Gill, Glover, Goetz, Goetz, Gomes, Goncharov, Gonz\'alez, Gonzalez~Castro, Gopakumar, Gorodetsky, Gossan, Gosselin, Gouaty, Grado, Graef, Granata, Grant, Gras, Gray, Greco, Green, Gretarsson, Groot, Grote, Grunewald, Gruning, Guidi, Guo, Gupta, Gupta, Gushwa, Gustafson, Gustafson, Halim, Hall, Hall, Hamilton, Hammond, Haney, Hanke, Hanks, Hanna, Hannam, Hannuksela, Hanson, Hardwick, Harms, Harry, Harry, Hart, Haster, Haughian, Healy, Heidmann, Heintze, Heitmann, Hello, Hemming, Hendry, Heng, Hennig, Heptonstall, Heurs, Hild, Hinderer, Ho, Hoak, Hofman, Holt, Holz, Hopkins, Horst, Hough, Houston, Howell,
  Hreibi, Hu, Huerta, Huet, Hughey, Husa, Huttner, Huynh-Dinh, Indik, Inta, Intini, Isa, Isac, Isi, Iyer, Izumi, Jacqmin, Jani, Jaranowski, Jawahar, Jim\'enez-Forteza, Johnson, Johnson-McDaniel, Jones, Jones, Jonker, Ju, Junker, Kalaghatgi, Kalogera, Kamai, Kandhasamy, Kang, Kanner, Kapadia, Karki, Karvinen, Kasprzack, Kastaun, Katolik, Katsavounidis, Katzman, Kaufer, Kawabe, K\'ef\'elian, Keitel, Kemball, Kennedy, Kent, Key, Khalili, Khan, Khan, Khan, Khazanov, Kijbunchoo, Kim, Kim, Kim, Kim, Kim, Kim, Kimbrell, King, King, Kinley-Hanlon, Kirchhoff, Kissel, Kleybolte, Klimenko, Knowles, Koch, Koehlenbeck, Koley, Kondrashov, Kontos, Korobko, Korth, Kowalska, Kozak, Kr\"amer, Kringel, Krishnan, Kr\'olak, Kuehn, Kumar, Kumar, Kumar, Kuo, Kutynia, Kwang, Lackey, Lai, Landry, Lang, Lange, Lantz, Lanza, Larson, Lartaux-Vollard, Lasky, Laxen, Lazzarini, Lazzaro, Leaci, Leavey, Lee, Lee, Lee, Lee, Lee, Lehmann, Lenon, Leon, Leonardi, Leroy, Letendre, Levin, Li, Linker, Littenberg, Liu, Liu, Lo, Lockerbie, London,
  Lord, Lorenzini, Loriette, Lormand, Losurdo, Lough, Lousto, Lovelace, L\"uck, Lumaca, Lundgren, Lynch, Ma, Macas, Macfoy, Machenschalk, MacInnis, Macleod, Maga\~na Hernandez, Maga\~na Sandoval, Maga\~na Zertuche, Magee, Majorana, Maksimovic, Man, Mandic, Mangano, Mansell, Manske, Mantovani, Marchesoni, Marion, M\'arka, M\'arka, Markakis, Markosyan, Markowitz, Maros, Marquina, Marsh, Martelli, Martellini, Martin, Martin, Martynov, Marx, Mason, Massera, Masserot, Massinger, Masso-Reid, Mastrogiovanni, Matas, Matichard, Matone, Mavalvala, Mazumder, McCarthy, McClelland, McCormick, McCuller, McGuire, McIntyre, McIver, McManus, McNeill, McRae, McWilliams, Meacher, Meadors, Mehmet, Meidam, Mejuto-Villa, Melatos, Mendell, Mercer, Merilh, Merzougui, Meshkov, Messenger, Messick, Metzdorff, Meyers, Miao, Michel, Middleton, Mikhailov, Milano, Miller, Miller, Miller, Millhouse, Milovich-Goff, Minazzoli, Minenkov, Ming, Mishra, Mitra, Mitrofanov, Mitselmakher, Mittleman, Moffa, Moggi, Mogushi, Mohan, Mohapatra, Molina,
  Montani, Moore, Moraru, Moreno, Morisaki, Morriss, Mours, Mow-Lowry, Mueller, Muir, Mukherjee, Mukherjee, Mukherjee, Mukund, Mullavey, Munch, Mu\~niz, Muratore, Murray, Nagar, Napier, Nardecchia, Naticchioni, Nayak, Neilson, Nelemans, Nelson, Nery, Neunzert, Nevin, Newport, Newton, Ng, Nguyen, Nguyen, Nichols, Nielsen, Nissanke, Nitz, Noack, Nocera, Nolting, North, Nuttall, Oberling, O'Dea, Ogin, Oh, Oh, Ohme, Okada, Oliver, Oppermann, Oram, O'Reilly, Ormiston, Ortega, O'Shaughnessy, Ossokine, Ottaway, Overmier, Owen, Pace, Page, Page, Pai, Pai, Palamos, Palashov, Palomba, Pal-Singh, Pan, Pan, Pang, Pang, Pankow, Pannarale, Pant, Paoletti, Paoli, Papa, Parida, Parker, Pascucci, Pasqualetti, Passaquieti, Passuello, Patil, Patricelli, Pearlstone, Pedraza, Pedurand, Pekowsky, Pele, Penn, Perez, Perreca, Perri, Pfeiffer, Phelps, Piccinni, Pichot, Piergiovanni, Pierro, Pillant, Pinard, Pinto, Pirello, Pitkin, Poe, Poggiani, Popolizio, Porter, Post, Powell, Prasad, Pratt, Pratten, Predoi, Prestegard, Prijatelj,
  Principe, Privitera, Prix, Prodi, Prokhorov, Puncken, Punturo, Puppo, P\"urrer, Qi, Quetschke, Quintero, Quitzow-James, Raab, Rabeling, Radkins, Raffai, Raja, Rajan, Rajbhandari, Rakhmanov, Ramirez, Ramos-Buades, Rapagnani, Raymond, Razzano, Read, Regimbau, Rei, Reid, Reitze, Ren, Reyes, Ricci, Ricker, Rieger, Riles, Rizzo, Robertson, Robie, Robinet, Rocchi, Rolland, Rollins, Roma, Romano, Romano, Romel, Romie, Rosi\ifmmode~\acute{n}\else \'{n}\fi{}ska, Ross, Rowan, R\"udiger, Ruggi, Rutins, Ryan, Sachdev, Sadecki, Sadeghian, Sakellariadou, Salconi, Saleem, Salemi, Samajdar, Sammut, Sampson, Sanchez, Sanchez, Sanchis-Gual, Sandberg, Sanders, Sassolas, Sathyaprakash, Saulson, Sauter, Savage, Sawadsky, Schale, Scheel, Scheuer, Schmidt, Schmidt, Schnabel, Schofield, Sch\"onbeck, Schreiber, Schuette, Schulte, Schutz, Schwalbe, Scott, Scott, Seidel, Sellers, Sengupta, Sentenac, Sequino, Sergeev, Shaddock, Shaffer, Shah, Shahriar, Shaner, Shao, Shapiro, Shawhan, Sheperd, Shoemaker, Shoemaker, Siellez, Siemens,
  Sieniawska, Sigg, Silva, Singer, Singh, Singhal, Sintes, Slagmolen, Smith, Smith, Smith, Somala, Son, Sonnenberg, Sorazu, Sorrentino, Souradeep, Spencer, Srivastava, Staats, Staley, Steinke, Steinlechner, Steinlechner, Steinmeyer, Stevenson, Stone, Stops, Strain, Stratta, Strigin, Strunk, Sturani, Stuver, Summerscales, Sun, Sunil, Suresh, Sutton, Swinkels, Szczepa\ifmmode~\acute{n}\else \'{n}\fi{}czyk, Tacca, Tait, Talbot, Talukder, Tanner, T\'apai, Taracchini, Tasson, Taylor, Taylor, Tewari, Theeg, Thies, Thomas, Thomas, Thomas, Thorne, Thorne, Thrane, Tiwari, Tiwari, Tokmakov, Toland, Tonelli, Tornasi, Torres-Forn\'e, Torrie, T\"oyr\"a, Travasso, Traylor, Trinastic, Tringali, Trozzo, Tsang, Tse, Tso, Tsukada, Tsuna, Tuyenbayev, Ueno, Ugolini, Unnikrishnan, Urban, Usman, Vahlbruch, Vajente, Valdes, Vallisneri, van Bakel, van Beuzekom, van~den Brand, Van Den~Broeck, Vander-Hyde, van~der Schaaf, van Heijningen, van Veggel, Vardaro, Varma, Vass, Vas\'uth, Vecchio, Vedovato, Veitch, Veitch, Venkateswara,
  Venugopalan, Verkindt, Vetrano, Vicer\'e, Viets, Vinciguerra, Vine, Vinet, Vitale, Vo, Vocca, Vorvick, Vyatchanin, Wade, Wade, Wade, Walet, Walker, Wallace, Walsh, Wang, Wang, Wang, Wang, Wang, Ward, Warner, Was, Watchi, Weaver, Wei, Weinert, Weinstein, Weiss, Wen, Wessel, We\ss{}els, Westerweck, Westphal, Wette, Whelan, Whitcomb, Whiting, Whittle, Wilken, Williams, Williams, Williamson, Willis, Willke, Wimmer, Winkler, Wipf, Wittel, Woan, Woehler, Wofford, Wong, Worden, Wright, Wu, Wysocki, Xiao, Yamamoto, Yancey, Yang, Yap, Yazback, Yu, Yu, Yvert, Zadro\ifmmode~\dot{z}\else \.{z}\fi{}ny, Zanolin, Zelenova, Zendri, Zevin, Zhang, Zhang, Zhang, Zhang, Zhao, Zhou, Zhou, Zhu, Zhu, Zimmerman, Zucker, \& Zweizig}]{gw170817}
Abbott, B.~P., Abbott, R., Abbott, T.~D., {et~al.} 2017{\natexlab{a}}, Phys. Rev. Lett., 119, 161101, \dodoi{10.1103/PhysRevLett.119.161101}

\bibitem[{Abbott {et~al.}(2017{\natexlab{b}})Abbott, Abbott, Abbott, Acernese, Ackley, Adams, Adams, Addesso, Adhikari, Adya, Affeldt, Afrough, Agarwal, Agathos, Agatsuma, Aggarwal, Aguiar, Aiello, Ain, Ajith, Allen, Allen, Allocca, Altin, Amato, Ananyeva, Anderson, Anderson, Angelova, Antier, Appert, Arai, Araya, Areeda, Arnaud, Arun, Ascenzi, Ashton, Ast, Aston, Astone, Atallah, Aufmuth, Aulbert, AultONeal, Austin, Avila-Alvarez, Babak, Bacon, Bader, Bae, Baker, Baldaccini, Ballardin, Ballmer, Banagiri, Barayoga, Barclay, Barish, Barker, Barkett, Barone, Barr, Barsotti, Barsuglia, Barta, Barthelmy, Bartlett, Bartos, Bassiri, Basti, Batch, Bawaj, Bayley, Bazzan, Bécsy, Beer, Bejger, Belahcene, Bell, Berger, Bergmann, Bero, Berry, Bersanetti, Bertolini, Betzwieser, Bhagwat, Bhandare, Bilenko, Billingsley, Billman, Birch, Birney, Birnholtz, Biscans, Biscoveanu, Bisht, Bitossi, Biwer, Bizouard, Blackburn, Blackman, Blair, Blair, Blair, Bloemen, Bock, Bode, Boer, Bogaert, Bohe, Bondu, Bonilla, Bonnand, Boom,
  Bork, Boschi, Bose, Bossie, Bouffanais, Bozzi, Bradaschia, Brady, Branchesi, Brau, Briant, Brillet, Brinkmann, Brisson, Brockill, Broida, Brooks, Brown, Brown, Brunett, Buchanan, Buikema, Bulik, Bulten, Buonanno, Buskulic, Buy, Byer, Cabero, Cadonati, Cagnoli, Cahillane, Bustillo, Callister, Calloni, Camp, Canepa, Canizares, Cannon, Cao, Cao, Capano, Capocasa, Carbognani, Caride, Carney, Diaz, Casentini, Caudill, Cavaglià, Cavalier, Cavalieri, Cella, Cepeda, Cerdá-Durán, Cerretani, Cesarini, Chamberlin, Chan, Chao, Charlton, Chase, Chassande-Mottin, Chatterjee, Chatziioannou, Cheeseboro, Chen, Chen, Chen, Cheng, Chia, Chincarini, Chiummo, Chmiel, Cho, Cho, Chow, Christensen, Chu, Chua, Chua, Chung, Chung, Ciani, Ciolfi, Cirelli, Cirone, Clara, Clark, Clearwater, Cleva, Cocchieri, Coccia, Cohadon, Cohen, Colla, Collette, Cominsky, Jr., Conti, Cooper, Corban, Corbitt, Cordero-Carrión, Corley, Cornish, Corsi, Cortese, Costa, Coughlin, Coughlin, Coulon, Countryman, Couvares, Covas, Cowan, Coward, Cowart,
  Coyne, Coyne, Creighton, Creighton, Cripe, Crowder, Cullen, Cumming, Cunningham, Cuoco, Canton, Dálya, Danilishin, D’Antonio, Danzmann, Dasgupta, Costa, Dattilo, Dave, Davier, Davis, Daw, Day, De, DeBra, Degallaix, Laurentis, Deléglise, Pozzo, Demos, Denker, Dent, Pietri, Dergachev, Rosa, DeRosa, Rossi, DeSalvo, de~Varona, Devenson, Dhurandhar, Díaz, Fiore, Giovanni, Girolamo, Lieto, Pace, Palma, Renzo, Doctor, Dolique, Donovan, Dooley, Doravari, Dorrington, Douglas, Álvarez, Downes, Drago, Dreissigacker, Driggers, Du, Ducrot, Dupej, Dwyer, Edo, Edwards, Effler, Ehrens, Eichholz, Eikenberry, Eisenstein, Essick, Estevez, Etienne, Etzel, Evans, Evans, Factourovich, Fafone, Fair, Fairhurst, Fan, Farinon, Farr, Farr, Fauchon-Jones, Favata, Fays, Fee, Fehrmann, Feicht, Fejer, Fernandez-Galiana, Ferrante, Ferreira, Ferrini, Fidecaro, Finstad, Fiori, Fiorucci, Fishbach, Fisher, Fitz-Axen, Flaminio, Fletcher, Fong, Font, Forsyth, Forsyth, Fournier, Frasca, Frasconi, Frei, Freise, Frey, Frey, Fries, Fritschel,
  Frolov, Fulda, Fyffe, Gabbard, Gadre, Gaebel, Gair, Gammaitoni, Ganija, Gaonkar, Garcia-Quiros, Garufi, Gateley, Gaudio, Gaur, Gayathri, Gehrels, Gemme, Genin, Gennai, George, George, Gergely, Germain, Ghonge, Ghosh, Ghosh, Ghosh, Giaime, Giardina, Giazotto, Gill, Glover, Goetz, Goetz, Gomes, Goncharov, González, Castro, Gopakumar, Gorodetsky, Gossan, Gosselin, Gouaty, Grado, Graef, Granata, Grant, Gras, Gray, Greco, Green, Gretarsson, Griswold, Groot, Grote, Grunewald, Gruning, Guidi, Guo, Gupta, Gupta, Gushwa, Gustafson, Gustafson, Halim, Hall, Hall, Hamilton, Hammond, Haney, Hanke, Hanks, Hanna, Hannam, Hannuksela, Hanson, Hardwick, Harms, Harry, Harry, Hart, Haster, Haughian, Healy, Heidmann, Heintze, Heitmann, Hello, Hemming, Hendry, Heng, Hennig, Heptonstall, Heurs, Hild, Hinderer, Hoak, Hofman, Holt, Holz, Hopkins, Horst, Hough, Houston, Howell, Hreibi, Hu, Huerta, Huet, Hughey, Husa, Huttner, Huynh-Dinh, Indik, Inta, Intini, Isa, Isac, Isi, Iyer, Izumi, Jacqmin, Jani, Jaranowski, Jawahar,
  Jiménez-Forteza, Johnson, Jones, Jones, Jonker, Ju, Junker, Kalaghatgi, Kalogera, Kamai, Kandhasamy, Kang, Kanner, Kapadia, Karki, Karvinen, Kasprzack, Katolik, Katsavounidis, Katzman, Kaufer, Kawabe, Kéfélian, Keitel, Kemball, Kennedy, Kent, Key, Khalili, Khan, Khan, Khan, Khazanov, Kijbunchoo, Kim, Kim, Kim, Kim, Kim, Kim, Kimbrell, King, King, Kinley-Hanlon, Kirchhoff, Kissel, Kleybolte, Klimenko, Knowles, Koch, Koehlenbeck, Koley, Kondrashov, Kontos, Korobko, Korth, Kowalska, Kozak, Krämer, Kringel, Krishnan, Królak, Kuehn, Kumar, Kumar, Kumar, Kuo, Kutynia, Kwang, Lackey, Lai, Landry, Lang, Lange, Lantz, Lanza, Larson, Lartaux-Vollard, Lasky, Laxen, Lazzarini, Lazzaro, Leaci, Leavey, Lee, Lee, Lee, Lee, Lee, Lehmann, Lenon, Leonardi, Leroy, Letendre, Levin, Li, Linker, Littenberg, Liu, Lo, Lockerbie, London, Lord, Lorenzini, Loriette, Lormand, Losurdo, Lough, Lousto, Lovelace, Lück, Lumaca, Lundgren, Lynch, Ma, Macas, Macfoy, Machenschalk, MacInnis, Macleod, Hernandez, Magaña-Sandoval, Zertuche,
  Magee, Majorana, Maksimovic, Man, Mandic, Mangano, Mansell, Manske, Mantovani, Marchesoni, Marion, Márka, Márka, Markakis, Markosyan, Markowitz, Maros, Marquina, Marsh, Martelli, Martellini, Martin, Martin, Martynov, Mason, Massera, Masserot, Massinger, Masso-Reid, Mastrogiovanni, Matas, Matichard, Matone, Mavalvala, Mazumder, McCarthy, McClelland, McCormick, McCuller, McGuire, McIntyre, McIver, McManus, McNeill, McRae, McWilliams, Meacher, Meadors, Mehmet, Meidam, Mejuto-Villa, Melatos, Mendell, Mercer, Merilh, Merzougui, Meshkov, Messenger, Messick, Metzdorff, Meyers, Miao, Michel, Middleton, Mikhailov, Milano, Miller, Miller, Miller, Millhouse, Milovich-Goff, Minazzoli, Minenkov, Ming, Mishra, Mitra, Mitrofanov, Mitselmakher, Mittleman, Moffa, Moggi, Mogushi, Mohan, Mohapatra, Montani, Moore, Moraru, Moreno, Morriss, Mours, Mow-Lowry, Mueller, Muir, Mukherjee, Mukherjee, Mukherjee, Mukund, Mullavey, Munch, Muñiz, Muratore, Murray, Napier, Nardecchia, Naticchioni, Nayak, Neilson, Nelemans, Nelson,
  Nery, Neunzert, Nevin, Newport, Newton, Ng, Nguyen, Nguyen, Nichols, Nielsen, Nissanke, Nitz, Noack, Nocera, Nolting, North, Nuttall, Oberling, O’Dea, Ogin, Oh, Oh, Ohme, Okada, Oliver, Oppermann, Oram, O’Reilly, Ormiston, Ortega, O’Shaughnessy, Ossokine, Ottaway, Overmier, Owen, Pace, Page, Page, Pai, Pai, Palamos, Palashov, Palomba, Pal-Singh, Pan, Pan, Pang, Pang, Pankow, Pannarale, Pant, Paoletti, Paoli, Papa, Parida, Parker, Pascucci, Pasqualetti, Passaquieti, Passuello, Patil, Patricelli, Pearlstone, Pedraza, Pedurand, Pekowsky, Pele, Penn, Perez, Perreca, Perri, Pfeiffer, Phelps, Piccinni, Pichot, Piergiovanni, Pierro, Pillant, Pinard, Pinto, Pirello, Pitkin, Poe, Poggiani, Popolizio, Porter, Post, Powell, Prasad, Pratt, Pratten, Predoi, Prestegard, Price, Prijatelj, Principe, Privitera, Prodi, Prokhorov, Puncken, Punturo, Puppo, Pürrer, Qi, Quetschke, Quintero, Quitzow-James, Raab, Rabeling, Radkins, Raffai, Raja, Rajan, Rajbhandari, Rakhmanov, Ramirez, Ramos-Buades, Rapagnani, Raymond,
  Razzano, Read, Regimbau, Rei, Reid, Reitze, Ren, Reyes, Ricci, Ricker, Rieger, Riles, Rizzo, Robertson, Robie, Robinet, Rocchi, Rolland, Rollins, Roma, Romano, Romel, Romie, Rosińska, Ross, Rowan, Rüdiger, Ruggi, Rutins, Ryan, Sachdev, Sadecki, Sadeghian, Sakellariadou, Salconi, Saleem, Salemi, Samajdar, Sammut, Sampson, Sanchez, Sanchez, Sanchis-Gual, Sandberg, Sanders, Sassolas, Sathyaprakash, Saulson, Sauter, Savage, Sawadsky, Schale, Scheel, Scheuer, Schmidt, Schmidt, Schnabel, Schofield, Schönbeck, Schreiber, Schuette, Schulte, Schutz, Schwalbe, Scott, Scott, Seidel, Sellers, Sengupta, Sentenac, Sequino, Sergeev, Shaddock, Shaffer, Shah, Shahriar, Shaner, Shao, Shapiro, Shawhan, Sheperd, Shoemaker, Shoemaker, Siellez, Siemens, Sieniawska, Sigg, Silva, Singer, Singh, Singhal, Sintes, Slagmolen, Smith, Smith, Smith, Somala, Son, Sonnenberg, Sorazu, Sorrentino, Souradeep, Spencer, Srivastava, Staats, Staley, Steinke, Steinlechner, Steinlechner, Steinmeyer, Stevenson, Stone, Stops, Strain, Stratta,
  Strigin, Strunk, Sturani, Stuver, Summerscales, Sun, Sunil, Suresh, Sutton, Swinkels, Szczepańczyk, Tacca, Tait, Talbot, Talukder, Tanner, Tápai, Taracchini, Tasson, Taylor, Taylor, Tewari, Theeg, Thies, Thomas, Thomas, Thomas, Thorne, Thorne, Thrane, Tiwari, Tiwari, Tokmakov, Toland, Tonelli, Tornasi, Torres-Forné, Torrie, Töyrä, Travasso, Traylor, Trinastic, Tringali, Trozzo, Tsang, Tse, Tso, Tsukada, Tsuna, Tuyenbayev, Ueno, Ugolini, Unnikrishnan, Urban, Usman, Vahlbruch, Vajente, Valdes, van Bakel, van Beuzekom, van~den Brand, Broeck, Vander-Hyde, van~der Schaaf, van Heijningen, van Veggel, Vardaro, Varma, Vass, Vasúth, Vecchio, Vedovato, Veitch, Veitch, Venkateswara, Venugopalan, Verkindt, Vetrano, Viceré, Viets, Vinciguerra, Vine, Vinet, Vitale, Vo, Vocca, Vorvick, Vyatchanin, Wade, Wade, Wade, Walet, Walker, Wallace, Walsh, Wang, Wang, Wang, Wang, Wang, Ward, Warner, Was, Watchi, Weaver, Wei, Weinert, Weinstein, Weiss, Wen, Wessel, Wessels, Westerweck, Westphal, Wette, Whelan, Whitcomb,
  Whiting, Whittle, Wilken, Williams, Williams, Williamson, Willis, Willke, Wimmer, Winkler, Wipf, Wittel, Woan, Woehler, Wofford, Wong, Worden, Wright, Wu, Wysocki, Xiao, Yamamoto, Yancey, Yang, Yap, Yazback, Yu, Yu, Yvert, Zadrożny, Zanolin, Zelenova, Zendri, Zevin, Zhang, Zhang, Zhang, Zhang, Zhao, Zhou, Zhou, Zhu, Zhu, Zimmerman, Zucker, Zweizig, Collaboration, Collaboration), Wilson-Hodge, Bissaldi, Blackburn, Briggs, Burns, Cleveland, Connaughton, Gibby, Giles, Goldstein, Hamburg, Jenke, Hui, Kippen, Kocevski, McBreen, Meegan, Paciesas, Poolakkil, Preece, Racusin, Roberts, Stanbro, Veres, von Kienlin, GBM), Savchenko, Ferrigno, Kuulkers, Bazzano, Bozzo, Brandt, Chenevez, Courvoisier, Diehl, Domingo, Hanlon, Jourdain, Laurent, Lebrun, Lutovinov, Martin-Carrillo, Mereghetti, Natalucci, Rodi, Roques, Sunyaev, Ubertini, (INTEGRAL), Aartsen, Ackermann, Adams, Aguilar, Ahlers, Ahrens, Samarai, Altmann, Andeen, Anderson, Ansseau, Anton, Argüelles, Auffenberg, Axani, Bagherpour, Bai, Barron, Barwick, Baum,
  Bay, Beatty, Tjus, Bernardini, Besson, Binder, Bindig, Blaufuss, Blot, Bohm, Börner, Bos, Bose, Böser, Botner, Bourbeau, Bourbeau, Bradascio, Braun, Brayeur, Brenzke, Bretz, Bron, Brostean-Kaiser, Burgman, Carver, Casey, Casier, Cheung, Chirkin, Christov, Clark, Classen, Coenders, Collin, Conrad, Cowen, Cross, Day, de~André, Clercq, DeLaunay, Dembinski, Ridder, Desiati, de~Vries, de~Wasseige, de~With, DeYoung, Díaz-Vélez, di~Lorenzo, Dujmovic, Dumm, Dunkman, Dvorak, Eberhardt, Ehrhardt, Eichmann, Eller, Evenson, Fahey, Fazely, Felde, Filimonov, Finley, Flis, Franckowiak, Friedman, Fuchs, Gaisser, Gallagher, Gerhardt, Ghorbani, Giang, Glauch, Glüsenkamp, Goldschmidt, Gonzalez, Grant, Griffith, Haack, Hallgren, Halzen, Hanson, Hebecker, Heereman, Helbing, Hellauer, Hickford, Hignight, Hill, Hoffman, Hoffmann, Hokanson-Fasig, Hoshina, Huang, Huber, Hultqvist, Hünnefeld, In, Ishihara, Jacobi, Japaridze, Jeong, Jero, Jones, Kalaczynski, Kang, Kappes, Karg, Karle, Kauer, Keivani, Kelley, Kheirandish, Kim,
  Kim, Kintscher, Kiryluk, Kittler, Klein, Kohnen, Koirala, Kolanoski, Köpke, Kopper, Kopper, Koschinsky, Koskinen, Kowalski, Krings, Kroll, Krückl, Kunnen, Kunwar, Kurahashi, Kuwabara, Kyriacou, Labare, Lanfranchi, Larson, Lauber, Lesiak-Bzdak, Leuermann, Liu, Lu, Lünemann, Luszczak, Madsen, Maggi, Mahn, Mancina, Maruyama, Mase, Maunu, McNally, Meagher, Medici, Meier, Menne, Merino, Meures, Miarecki, Micallef, Momenté, Montaruli, Moore, Moulai, Nahnhauer, Nakarmi, Naumann, Neer, Niederhausen, Nowicki, Nygren, Pollmann, Olivas, O’Murchadha, Palczewski, Pandya, Pankova, Peiffer, Pepper, de~los Heros, Pieloth, Pinat, Price, Przybylski, Raab, Rädel, Rameez, Rawlins, Rea, Reimann, Relethford, Relich, Resconi, Rhode, Richman, Robertson, Rongen, Rott, Ruhe, Ryckbosch, Rysewyk, Sälzer, Herrera, Sandrock, Sandroos, Santander, Sarkar, Sarkar, Satalecka, Schlunder, Schmidt, Schneider, Schoenen, Schöneberg, Schumacher, Seckel, Seunarine, Soedingrekso, Soldin, Song, Spiczak, Spiering, Stachurska, Stamatikos,
  Stanev, Stasik, Stettner, Steuer, Stezelberger, Stokstad, Stössl, Strotjohann, Stuttard, Sullivan, Sutherland, Taboada, Tatar, Tenholt, Ter-Antonyan, Terliuk, Tešić, Tilav, Toale, Tobin, Toscano, Tosi, Tselengidou, Tung, Turcati, Turley, Ty, Unger, Usner, Vandenbroucke, Driessche, van Eijndhoven, Vanheule, van Santen, Vehring, Vogel, Vraeghe, Walck, Wallace, Wallraff, Wandler, Wandkowsky, Waza, Weaver, Weiss, Wendt, Werthebach, Whelan, Wiebe, Wiebusch, Wille, Williams, Wills, Wolf, Wood, Woolsey, Woschnagg, Xu, Xu, Xu, Yanez, Yodh, Yoshida, Yuan, Zoll, Collaboration), Balasubramanian, Mate, Bhalerao, Bhattacharya, Vibhute, Dewangan, Rao, Vadawale, Team), Svinkin, Hurley, Aptekar, Frederiks, Golenetskii, Kozlova, Lysenko, Oleynik, Tsvetkova, Ulanov, Cline, Collaboration), Li, Xiong, Zhang, Lu, Song, Cao, Chang, Chen, Chen, Chen, Chen, Chen, Chen, Cui, Cui, Deng, Dong, Du, Fu, Gao, Gao, Gao, Ge, Gu, Guan, Guo, Han, Hu, Huang, Huo, Jia, Jiang, Jiang, Jin, Jin, Li, Li, Li, Li, Li, Li, Li, Li, Li, Li, Li,
  Liang, Liao, Liu, Liu, Liu, Liu, Liu, Liu, Liu, Lu, Lu, Luo, Ma, Meng, Nang, Nie, Ou, Qu, Sai, Sun, Tan, Tao, Tao, Tuo, Wang, Wang, Wang, Wang, Wang, Wen, Wu, Wu, Xiao, Xu, Xu, Yan, Yang, Yang, Yang, Zhang, Zhang, Zhang, Zhang, Zhang, Zhang, Zhang, Zhang, Zhang, Zhang, Zhang, Zhang, Zhang, Zhang, Zhang, Zhang, Zhang, Zhang, Zhao, Zhao, Zhao, Zheng, Zhu, Zhu, Zou, Collaboration), Albert, André, Anghinolfi, Ardid, Aubert, Aublin, Avgitas, Baret, Barrios-Martí, Basa, Belhorma, Bertin, Biagi, Bormuth, Bourret, Bouwhuis, Brânzaş, Bruijn, Brunner, Busto, Capone, Caramete, Carr, Celli, Moursli, Chiarusi, Circella, Coelho, Coleiro, Coniglione, Costantini, Coyle, Creusot, Díaz, Deschamps, Bonis, Distefano, Palma, Domi, Donzaud, Dornic, Drouhin, Eberl, Bojaddaini, Khayati, Elsässer, Enzenhöfer, Ettahiri, Fassi, Felis, Fusco, Gay, Giordano, Glotin, Grégoire, Ruiz, Graf, Hallmann, van Haren, Heijboer, Hello, Hernández-Rey, Hössl, Hofestädt, Hugon, Illuminati, James, de~Jong, Jongen, Kadler, Kalekin, Katz,
  Kiessling, Kouchner, Kreter, Kreykenbohm, Kulikovskiy, Lachaud, Lahmann, Lefèvre, Leonora, Lotze, Loucatos, Marcelin, Margiotta, Marinelli, Martínez-Mora, Mele, Melis, Michael, Migliozzi, Moussa, Navas, Nezri, Organokov, Păvălaş, Pellegrino, Perrina, Piattelli, Popa, Pradier, Quinn, Racca, Riccobene, Sánchez-Losa, Saldaña, Salvadori, Samtleben, Sanguineti, Sapienza, Sieger, Spurio, Stolarczyk, Taiuti, Tayalati, Trovato, Turpin, Tönnis, Vallage, Elewyck, Versari, Vivolo, Vizzoca, Wilms, Zornoza, Zúñiga, Collaboration), Beardmore, Breeveld, Burrows, Cenko, Cusumano, D’Aì, de~Pasquale, Emery, Evans, Giommi, Gronwall, Kennea, Krimm, Kuin, Lien, Marshall, Melandri, Nousek, Oates, Osborne, Pagani, Page, Palmer, Perri, Siegel, Sbarufatti, Tagliaferri, Tohuvavohu, Collaboration), Tavani, Verrecchia, Bulgarelli, Evangelista, Pacciani, Feroci, Pittori, Giuliani, Monte, Donnarumma, Argan, Trois, Ursi, Cardillo, Piano, Longo, Lucarelli, Munar-Adrover, Fuschino, Labanti, Marisaldi, Minervini, Fioretti,
  Parmiggiani, Gianotti, Trifoglio, Persio, Antonelli, Barbiellini, Caraveo, Cattaneo, Costa, Colafrancesco, D’Amico, Ferrari, Morselli, Paoletti, Picozza, Pilia, Rappoldi, Soffitta, Vercellone, Team), Foley, Coulter, Kilpatrick, Drout, Piro, Shappee, Siebert, Simon, Ulloa, Kasen, Madore, Murguia-Berthier, Pan, Prochaska, Ramirez-Ruiz, Rest, Rojas-Bravo, Team), Berger, Soares-Santos, Annis, Alexander, Allam, Balbinot, Blanchard, Brout, Butler, Chornock, Cook, Cowperthwaite, Diehl, Drlica-Wagner, Drout, Durret, Eftekhari, Finley, Fong, Frieman, Fryer, García-Bellido, Gruendl, Hartley, Herner, Kessler, Lin, Lopes, Lourenço, Margutti, Marshall, Matheson, Medina, Metzger, Muñoz, Muir, Nicholl, Nugent, Palmese, Paz-Chinchón, Quataert, Sako, Sauseda, Schlegel, Scolnic, Secco, Smith, Sobreira, Villar, Vivas, Wester, Williams, Yanny, Zenteno, Zhang, Abbott, Banerji, Bechtol, Benoit-Lévy, Bertin, Brooks, Buckley-Geer, Burke, Capozzi, Rosell, Kind, Castander, Crocce, Cunha, D’Andrea, da~Costa, Davis, DePoy,
  Desai, Dietrich, Eifler, Fernandez, Flaugher, Fosalba, Gaztanaga, Gerdes, Giannantonio, Goldstein, Gruen, Gschwend, Gutierrez, Honscheid, James, Jeltema, Johnson, Johnson, Kent, Krause, Kron, Kuehn, Lahav, Lima, Maia, March, Martini, McMahon, Menanteau, Miller, Miquel, Mohr, Nichol, Ogando, Plazas, Romer, Roodman, Rykoff, Sanchez, Scarpine, Schindler, Schubnell, Sevilla-Noarbe, Sheldon, Smith, Smith, Stebbins, Suchyta, Swanson, Tarle, Thomas, Troxel, Tucker, Vikram, Walker, Wechsler, Weller, Carlin, Gill, Li, Marriner, Neilsen, Collaboration, the DES~Collaboration), Haislip, Kouprianov, Reichart, Sand, Tartaglia, Valenti, Yang, Collaboration), Benetti, Brocato, Campana, Cappellaro, Covino, D’Avanzo, D’Elia, Getman, Ghirlanda, Ghisellini, Limatola, Nicastro, Palazzi, Pian, Piranomonte, Possenti, Rossi, Salafia, Tomasella, Amati, Antonelli, Bernardini, Bufano, Capaccioli, Casella, Dadina, Cesare, Paola, Giuffrida, Giunta, Israel, Lisi, Maiorano, Mapelli, Masetti, Pescalli, Pulone, Salvaterra, Schipani,
  Spera, Stamerra, Stella, Testa, Turatto, Vergani, Aresu, Bachetti, Buffa, Burgay, Buttu, Caria, Carretti, Casasola, Castangia, Carboni, Casu, Concu, Corongiu, Deiana, Egron, Fara, Gaudiomonte, Gusai, Ladu, Loru, Leurini, Marongiu, Melis, Melis, Migoni, Milia, Navarrini, Orlati, Ortu, Palmas, Pellizzoni, Perrodin, Pisanu, Poppi, Righini, Saba, Serra, Serrau, Stagni, Surcis, Vacca, Vargiu, Hunt, Jin, Klose, Kouveliotou, Mazzali, Møller, Nava, Piran, Selsing, Vergani, Wiersema, Toma, Higgins, Mundell, di~Serego~Alighieri, Gótz, Gao, Gomboc, Kaper, Kobayashi, Kopac, Mao, Starling, Steele, van~der Horst, TeAm), Acero, Atwood, Baldini, Barbiellini, Bastieri, Berenji, Bellazzini, Bissaldi, Blandford, Bloom, Bonino, Bottacini, Bregeon, Buehler, Buson, Cameron, Caputo, Caraveo, Cavazzuti, Chekhtman, Cheung, Chiang, Ciprini, Cohen-Tanugi, Cominsky, Costantin, Cuoco, D’Ammando, de~Palma, Digel, Lalla, Mauro, Venere, Dubois, Fegan, Focke, Franckowiak, Fukazawa, Funk, Fusco, Gargano, Gasparrini, Giglietto, Giordano,
  Giroletti, Glanzman, Green, Grondin, Guillemot, Guiriec, Harding, Horan, Jóhannesson, Kamae, Kensei, Kuss, Mura, Latronico, Lemoine-Goumard, Longo, Loparco, Lovellette, Lubrano, Magill, Maldera, Manfreda, Mazziotta, McEnery, Meyer, Michelson, Mirabal, Monzani, Moretti, Morselli, Moskalenko, Negro, Nuss, Ojha, Omodei, Orienti, Orlando, Palatiello, Paliya, Paneque, Pesce-Rollins, Piron, Porter, Principe, Rainò, Rando, Razzano, Razzaque, Reimer, Reimer, Reposeur, Rochester, Parkinson, Sgrò, Siskind, Spada, Spandre, Suson, Takahashi, Tanaka, Thayer, Thayer, Thompson, Tibaldo, Torres, Torresi, Troja, Venters, Vianello, Zaharijas, Collaboration), Allison, Bannister, Dobie, Kaplan, Lenc, Lynch, Murphy, Sadler, Array), Hotan, James, Oslowski, Raja, Shannon, Whiting, Pathfinder), Arcavi, Howell, McCully, Hosseinzadeh, Hiramatsu, Poznanski, Barnes, Zaltzman, Vasylyev, Maoz, Group), Cooke, Bailes, Wolf, Deller, Lidman, Wang, Gendre, Andreoni, Ackley, Pritchard, Bessell, Chang, Möller, Onken, Scalzo, Ridden-Harper,
  Sharp, Tucker, Farrell, Elmer, Johnston, Krishnan, Keane, Green, Jameson, Hu, Ma, Sun, Wu, Wang, Shang, Hu, Ashley, Yuan, Li, Tao, Zhu, Zhang, Suntzeff, Zhou, Yang, Orange, Morris, Cucchiara, Giblin, Klotz, Staff, Thierry, Schmidt, program), AST3, Collaborations), Tanvir, Levan, Cano, de~Ugarte-Postigo, González-Fernández, Greiner, Hjorth, Irwin, Krühler, Mandel, Milvang-Jensen, O’Brien, Rol, Rosetti, Rosswog, Rowlinson, Steeghs, Thöne, Ulaczyk, Watson, Bruun, Cutter, Jaimes, Fujii, Fruchter, Gompertz, Jakobsson, Hodosan, Jèrgensen, Kangas, Kann, Rabus, Schrøder, Stanway, Wijers, Collaboration), Lipunov, Gorbovskoy, Kornilov, Tyurina, Balanutsa, Kuznetsov, Vlasenko, Podesta, Lopez, Podesta, Levato, Saffe, Mallamaci, Budnev, Gress, Kuvshinov, Gorbunov, Vladimirov, Zimnukhov, Gabovich, Yurkov, Sergienko, Rebolo, Serra-Ricart, Tlatov, Ishmuhametova, Collaboration), Abe, Aoki, Aoki, Asakura, Baar, Barway, Bond, Doi, Finet, Fujiyoshi, Furusawa, Honda, Itoh, Kanda, Kawabata, Kawabata, Kim, Koshida,
  Kuroda, Lee, Liu, Matsubayashi, Miyazaki, Morihana, Morokuma, Motohara, Murata, Nagai, Nagashima, Nagayama, Nakaoka, Nakata, Ohsawa, Ohshima, Ohta, Okita, Saito, Saito, Sako, Sekiguchi, Sumi, Tajitsu, Takahashi, Takayama, Tamura, Tanaka, Tanaka, Terai, Tominaga, Tristram, Uemura, Utsumi, Yamaguchi, Yasuda, Yoshida, Zenko, (J-GEM), Adams, Anupama, Bally, Barway, Bellm, Blagorodnova, Cannella, Chandra, Chatterjee, Clarke, Cobb, Cook, Copperwheat, De, Emery, Feindt, Foster, Fox, Frail, Fremling, Frohmaier, Garcia, Ghosh, Giacintucci, Goobar, Gottlieb, Grefenstette, Hallinan, Harrison, Heida, Helou, Ho, Horesh, Hotokezaka, Ip, Itoh, Jacobs, Jencson, Kasen, Kasliwal, Kassim, Kim, Kiran, Kuin, Kulkarni, Kupfer, Lau, Madsen, Mazzali, Miller, Miyasaka, Mooley, Myers, Nakar, Ngeow, Nugent, Ofek, Palliyaguru, Pavana, Perley, Peters, Pike, Piran, Qi, Quimby, Rana, Rosswog, Rusu, Sadler, Sistine, Sollerman, Xu, Yan, Yatsu, Yu, Zhang, Zhao, TTU-NRAO, Collaborations), Chambers, Huber, Schultz, Bulger, Flewelling,
  Magnier, Lowe, Wainscoat, Waters, Willman, (Pan-STARRS), Ebisawa, Hanyu, Harita, Hashimoto, Hidaka, Hori, Ishikawa, Isobe, Iwakiri, Kawai, Kawai, Kawamuro, Kawase, Kitaoka, Makishima, Matsuoka, Mihara, Morita, Morita, Nakahira, Nakajima, Nakamura, Negoro, Oda, Sakamaki, Sasaki, Serino, Shidatsu, Shimomukai, Sugawara, Sugita, Sugizaki, Tachibana, Takao, Tanimoto, Tomida, Tsuboi, Tsunemi, Ueda, Ueno, Yamada, Yamaoka, Yamauchi, Yatabe, Yoneyama, Yoshii, Team), Coward, Crisp, Macpherson, Andreoni, Laugier, Noysena, Klotz, Gendre, Thierry, Turpin, Consortium), Im, Choi, Kim, Yoon, Lim, Lee, Lee, Kim, Ko, Joe, Kwon, Kim, Lim, Choi, Collaboration), Fynbo, Malesani, Xu, Telescope), Smartt, Jerkstrand, Kankare, Sim, Fraser, Inserra, Maguire, Leloudas, Magee, Shingles, Smith, Young, Kotak, Gal-Yam, Lyman, Homan, Agliozzo, Anderson, Angus, Ashall, Barbarino, Bauer, Berton, Botticella, Bulla, Cannizzaro, Cartier, Cikota, Clark, Cia, Valle, Dennefeld, Dessart, Dimitriadis, Elias-Rosa, Firth, Flörs, Frohmaier, Galbany,
  González-Gaitán, Gromadzki, Gutiérrez, Hamanowicz, Harmanen, Heintz, Hernandez, Hodgkin, Hook, Izzo, James, Jonker, Kerzendorf, Kostrzewa-Rutkowska, Kromer, Kuncarayakti, Lawrence, Manulis, Mattila, McBrien, Müller, Nordin, O’Neill, Onori, Palmerio, Pastorello, Patat, Pignata, Podsiadlowski, Razza, Reynolds, Roy, Ruiter, Rybicki, Salmon, Pumo, Prentice, Seitenzahl, Smith, Sollerman, Sullivan, Szegedi, Taddia, Taubenberger, Terreran, Soelen, Vos, Walton, Wright, Wyrzykowski, Yaron, (ePESSTO), Chen, Krühler, Schady, Wiseman, Greiner, Rau, Schweyer, Klose, Guelbenzu, (GROND), Palliyaguru, University), Shara, Williams, Vaisanen, Potter, Colmenero, Crawford, Buckley, Mao, Group), Díaz, Macri, Lambas, de~Oliveira, Castellón, Ribeiro, Sánchez, Schoenell, Abramo, Akras, Alcaniz, Artola, Beroiz, Bonoli, Cabral, Camuccio, Chavushyan, Coelho, Colazo, Costa-Duarte, Larenas, Romero, Dultzin, Fernández, García, Girardini, Gonçalves, Gonçalves, Gurovich, Jiménez-Teja, Kanaan, Lares, de~Oliveira,
  López-Cruz, Melia, Molino, Padilla, Peñuela, Placco, Quiñones, Rivera, Renzi, Riguccini, Ríos-López, Rodriguez, Sampedro, Schneiter, Sodré, Starck, Torres-Flores, Tornatore, Zadrożny, Castillo, of~the South~Collaboration), Castro-Tirado, Tello, Hu, Zhang, Cunniffe, Castellón, Hiriart, Caballero-García, Jelínek, Kubánek, del Pulgar, Park, Jeong, Cerón, Pandey, Yock, Querel, Fan, Wang, Collaboration), Beardsley, Brown, Crosse, Emrich, Franzen, Gaensler, Horsley, Johnston-Hollitt, Kenney, Morales, Pallot, Sokolowski, Steele, Tingay, Trott, Walker, Wayth, Williams, Wu, Array), Yoshida, Sakamoto, Kawakubo, Yamaoka, Takahashi, Asaoka, Ozawa, Torii, Shimizu, Tamura, Ishizaki, Cherry, Ricciarini, Penacchioni, Marrocchesi, Collaboration), Pozanenko, Volnova, Mazaeva, Minaev, Krugov, Kusakin, Reva, Moskvitin, Rumyantsev, Inasaridze, Klunko, Tungalag, Schmalz, Burhonov, up~Collaboration), Abdalla, Abramowski, Aharonian, Benkhali, Angüner, Arakawa, Arrieta, Aubert, Backes, Balzer, Barnard, Becherini, Tjus,
  Berge, Bernhard, Bernlöhr, Blackwell, Böttcher, Boisson, Bolmont, Bonnefoy, Bordas, Bregeon, Brun, Brun, Bryan, Büchele, Bulik, Capasso, Caroff, Carosi, Casanova, Cerruti, Chakraborty, Chaves, Chen, Chevalier, Colafrancesco, Condon, Conrad, Davids, Decock, Deil, Devin, deWilt, Dirson, Djannati-Ataï, Donath, Drury, Dutson, Dyks, Edwards, Egberts, Emery, Ernenwein, Eschbach, Farnier, Fegan, Fernandes, Fiasson, Fontaine, Funk, Füssling, Gabici, Gallant, Garrigoux, Gaté, Giavitto, Giebels, Glawion, Glicenstein, Gottschall, Grondin, Hahn, Haupt, Hawkes, Heinzelmann, Henri, Hermann, Hinton, Hofmann, Hoischen, Holch, Holler, Horns, Ivascenko, Iwasaki, Jacholkowska, Jamrozy, Jankowsky, Jankowsky, Jingo, Jouvin, Jung-Richardt, Kastendieck, Katarzyński, Katsuragawa, Kerszberg, Khangulyan, Khélifi, King, Klepser, Klochkov, Kluźniak, Komin, Kosack, Krakau, Kraus, Krüger, Laffon, Lamanna, Lau, Lees, Lefaucheur, Lemière, Lemoine-Goumard, Lenain, Leser, Lohse, Lorentz, Liu, Lypova, Malyshev, Marandon,
  Marcowith, Mariaud, Marx, Maurin, Maxted, Mayer, Meintjes, Meyer, Mitchell, Moderski, Mohamed, Mohrmann, Morå, Moulin, Murach, Nakashima, de~Naurois, Ndiyavala, Niederwanger, Niemiec, Oakes, O’Brien, Odaka, Ohm, Ostrowski, Oya, Padovani, Panter, Parsons, Pekeur, Pelletier, Perennes, Petrucci, Peyaud, Piel, Pita, Poireau, Poon, Prokhorov, Prokoph, Pühlhofer, Punch, Quirrenbach, Raab, Rauth, Reimer, Reimer, Renaud, de~los Reyes, Rieger, Rinchiuso, Romoli, Rowell, Rudak, Rulten, Sahakian, Saito, Sanchez, Santangelo, Sasaki, Schlickeiser, Schüssler, Schulz, Schwanke, Schwemmer, Seglar-Arroyo, Settimo, Seyffert, Shafi, Shilon, Shiningayamwe, Simoni, Sol, Spanier, Spir-Jacob, Stawarz, Steenkamp, Stegmann, Steppa, Sushch, Takahashi, Tavernet, Tavernier, Taylor, Terrier, Tibaldo, Tiziani, Tluczykont, Trichard, Tsirou, Tsuji, Tuffs, Uchiyama, van~der Walt, van Eldik, van Rensburg, van Soelen, Vasileiadis, Veh, Venter, Viana, Vincent, Vink, Voisin, Völk, Vuillaume, Wadiasingh, Wagner, Wagner, Wagner, White,
  Wierzcholska, Willmann, Wörnlein, Wouters, Yang, Zaborov, Zacharias, Zanin, Zdziarski, Zech, Zefi, Ziegler, Zorn, Żywucka, Collaboration), Fender, Broderick, Rowlinson, Wijers, Stewart, ter Veen, Shulevski, Collaboration), Kavic, Simonetti, League, Tsai, Obenberger, Nathaniel, Taylor, Dowell, Liebling, Estes, Lippert, Sharma, Vincent, Farella, Array), Abeysekara, Albert, Alfaro, Alvarez, Arceo, Arteaga-Velázquez, Rojas, Solares, Barber, Gonzalez, Becerril, Belmont-Moreno, BenZvi, Berley, Bernal, Braun, Brisbois, Caballero-Mora, Capistrán, Carramiñana, Casanova, Castillo, Cotti, Cotzomi, de~León, León, la~Fuente, Hernandez, Dichiara, Dingus, DuVernois, Díaz-Vélez, Ellsworth, Engel, Enríquez-Rivera, Fiorino, Fleischhack, Fraija, García-González, Garfias, Gerhardt, Muñoz, González, Goodman, Hampel-Arias, Harding, Hernandez, Hernandez-Almada, Hona, Hüntemeyer, Iriarte, Jardin-Blicq, Joshi, Kaufmann, Kieda, Lara, Lauer, Lennarz, Vargas, Linnemann, Longinotti, Raya, Luna-García, López-Coto,
  Malone, Marinelli, Martinez, Martinez-Castellanos, Martínez-Castro, Martínez-Huerta, Matthews, Miranda-Romagnoli, Moreno, Mostafá, Nellen, Newbold, Nisa, Noriega-Papaqui, Pelayo, Pretz, Pérez-Pérez, Ren, Rho, Rivière, Rosa-González, Rosenberg, Ruiz-Velasco, Salazar, Greus, Sandoval, Schneider, Schoorlemmer, Sinnis, Smith, Springer, Surajbali, Tibolla, Tollefson, Torres, Ukwatta, Weisgarber, Westerhoff, Wisher, Wood, Yapici, Yodh, Younk, Zhou, Álvarez, Collaboration), Aab, Abreu, Aglietta, Albuquerque, Albury, Allekotte, Almela, Castillo, Alvarez-Muñiz, Anastasi, Anchordoqui, Andrada, Andringa, Aramo, Arsene, Asorey, Assis, Avila, Badescu, Balaceanu, Barbato, Luz, Becker, Bellido, Berat, Bertaina, Bertou, Biermann, Biteau, Blaess, Blanco, Blazek, Bleve, Boháčová, Bonifazi, Borodai, Botti, Brack, Brancus, Bretz, Bridgeman, Briechle, Buchholz, Bueno, Buitink, Buscemi, Caballero-Mora, Caccianiga, Cancio, Canfora, Caruso, Castellina, Catalani, Cataldi, Cazon, Chavez, Chinellato, Chudoba, Clay,
  Cerutti, Colalillo, Coleman, Collica, Coluccia, Conceição, Consolati, Contreras, Cooper, Coutu, Covault, Cronin, D’Amico, Daniel, Dasso, Daumiller, Dawson, Day, de~Almeida, de~Jong, Mauro, de~Mello~Neto, Mitri, de~Oliveira, de~Souza, Debatin, Deligny, Castro, Diogo, Dobrigkeit, D’Olivo, Dorosti, Anjos, Dova, Dundovic, Ebr, Engel, Erdmann, Erfani, Escobar, Espadanal, Etchegoyen, Falcke, Farmer, Farrar, Fauth, Fazzini, Feldbusch, Fenu, Fick, Figueira, Filipčič, Freire, Fujii, Fuster, Gaïor, García, Gaté, Gemmeke, Gherghel-Lascu, Ghia, Giaccari, Giammarchi, Giller, Głas, Glaser, Golup, Berisso, Vitale, González, Gorgi, Gottowik, Grillo, Grubb, Guarino, Guedes, Halliday, Hampel, Hansen, Harari, Harrison, Harvey, Haungs, Hebbeker, Heck, Heimann, Herve, Hill, Hojvat, Holt, Homola, Hörandel, Horvath, Hrabovský, Huege, Hulsman, Insolia, Isar, Jandt, Johnsen, Josebachuili, Jurysek, Kääpä, Kampert, Keilhauer, Kemmerich, Kemp, Kieckhafer, Klages, Kleifges, Kleinfeller, Krause, Krohm, Kuempel, Mezek,
  Kunka, Awad, Lago, LaHurd, Lang, Lauscher, Legumina, de~Oliveira, Letessier-Selvon, Lhenry-Yvon, Link, Presti, Lopes, López, Casado, Lorek, Luce, Lucero, Malacari, Mallamaci, Mandat, Mantsch, Mariazzi, Maris, Marsella, Martello, Martinez, Bravo, Meza, Mathes, Mathys, Matthews, Matthiae, Mayotte, Mazur, Medina, Medina-Tanco, Melo, Menshikov, Merenda, Michal, Micheletti, Middendorf, Miramonti, Mitrica, Mockler, Mollerach, Montanet, Morello, Morlino, Müller, Müller, Muller, Müller, Mussa, Naranjo, Nguyen, Niculescu-Oglinzanu, Niechciol, Niemietz, Niggemann, Nitz, Nosek, Novotny, Nožka, Núñez, Oikonomou, Olinto, Palatka, Pallotta, Papenbreer, Parente, Parra, Paul, Pech, Pedreira, Pȩkala, Peña-Rodriguez, Pereira, Perlin, Perrone, Peters, Petrera, Phuntsok, Pierog, Pimenta, Pirronello, Platino, Plum, Poh, Porowski, Prado, Privitera, Prouza, Quel, Querchfeld, Quinn, Ramos-Pollan, Rautenberg, Ravignani, Ridky, Riehn, Risse, Ristori, Rizi, de~Carvalho, Fernandez, Rojo, Roncoroni, Roth, Roulet, Rovero,
  Ruehl, Saffi, Saftoiu, Salamida, Salazar, Saleh, Salina, Sánchez, Sanchez-Lucas, Santos, Santos, Sarazin, Sarmento, Sarmiento-Cano, Sato, Schauer, Scherini, Schieler, Schimp, Schmidt, Scholten, Schovánek, Schröder, Schröder, Schulz, Schumacher, Sciutto, Segreto, Shadkam, Shellard, Sigl, Silli, Šmída, Snow, Sommers, Sonntag, Soriano, Squartini, Stanca, Stanič, Stasielak, Stassi, Stolpovskiy, Strafella, Streich, Suarez, Suarez-Durán, Sudholz, Suomijärvi, Supanitsky, Šupík, Swain, Szadkowski, Taboada, Taborda, Timmermans, Peixoto, Tomankova, Tomé, Elipe, Travnicek, Trini, Tueros, Ulrich, Unger, Urban, Galicia, Valiño, Valore, van Aar, van Bodegom, van~den Berg, van Vliet, Varela, Cárdenas, Vázquez, Veberič, Ventura, Quispe, Verzi, Vicha, Villaseñor, Vorobiov, Wahlberg, Wainberg, Walz, Watson, Weber, Weindl, Wiedeński, Wiencke, Wilczyński, Wirtz, Wittkowski, Wundheiler, Yang, Yushkov, Zas, Zavrtanik, Zavrtanik, Zepeda, Zimmermann, Ziolkowski, Zong, Zuccarello, Collaboration), Kim, Schulze,
  Bauer, Corral-Santana, de~Gregorio-Monsalvo, González-López, Hartmann, Ishwara-Chandra, Martín, Mehner, Misra, Michałowski, Resmi, Collaboration), Paragi, Agudo, An, Beswick, Casadio, Frey, Jonker, Kettenis, Marcote, Moldon, Szomoru, van Langevelde, Yang, Team), Cwiek, Cwiok, Czyrkowski, Dabrowski, Kasprowicz, Mankiewicz, Nawrocki, Opiela, Piotrowski, Wrochna, Zaremba, Żarnecki, of~the Sky~Collaboration), Haggard, Nynka, Ruan, at~McGill~University), Bland, Booler, Devillepoix, de~Gois, Hancock, Howie, Paxman, Sansom, Towner, Network), Tonry, Coughlin, Stubbs, Denneau, Heinze, Stalder, Weiland, (ATLAS), Eatough, Kramer, Kraus, Survey), Troja, Piro, González, Butler, Fox, Khandrika, Kutyrev, Lee, Ricci, Jr., Sánchez-Ramírez, Veilleux, Watson, Wieringa, Burgess, van Eerten, Fontes, Fryer, Korobkin, Wollaeger, (RIMAS, RATIR), Camilo, Foley, Goedhart, Makhathini, Oozeer, Smirnov, Fender, Woudt, \& Africa/MeerKAT)}]{Abbott_2017}
---. 2017{\natexlab{b}}, The Astrophysical Journal Letters, 848, L12, \dodoi{10.3847/2041-8213/aa91c9}

\bibitem[{Abbott {et~al.}(2019)Abbott, Abbott, Abbott, Acernese, Ackley, Adams, Adams, Addesso, Adhikari, Adya, Affeldt, Agarwal, Agathos, Agatsuma, Aggarwal, Aguiar, Aiello, Ain, Ajith, Allen, Allen, Allocca, Aloy, Altin, Amato, Ananyeva, Anderson, Anderson, Angelova, Antier, Appert, Arai, Araya, Areeda, Ar\`ene, Arnaud, Arun, Ascenzi, Ashton, Ast, Aston, Astone, Atallah, Aubin, Aufmuth, Aulbert, AultONeal, Austin, Avila-Alvarez, Babak, Bacon, Badaracco, Bader, Bae, Baker, Baldaccini, Ballardin, Ballmer, Banagiri, Barayoga, Barclay, Barish, Barker, Barkett, Barnum, Barone, Barr, Barsotti, Barsuglia, Barta, Bartlett, Bartos, Bassiri, Basti, Batch, Bawaj, Bayley, Bazzan, B\'ecsy, Beer, Bejger, Belahcene, Bell, Beniwal, Bensch, Berger, Bergmann, Bernuzzi, Bero, Berry, Bersanetti, Bertolini, Betzwieser, Bhandare, Bilenko, Bilgili, Billingsley, Billman, Birch, Birney, Birnholtz, Biscans, Biscoveanu, Bisht, Bitossi, Bizouard, Blackburn, Blackman, Blair, Blair, Blair, Bloemen, Bock, Bode, Boer, Boetzel, Bogaert,
  Bohe, Bondu, Bonilla, Bonnand, Booker, Boom, Booth, Bork, Boschi, Bose, Bossie, Bossilkov, Bosveld, Bouffanais, Bozzi, Bradaschia, Brady, Bramley, Branchesi, Brau, Briant, Brighenti, Brillet, Brinkmann, Brisson, Brockill, Brooks, Brown, Brunett, Buchanan, Buikema, Bulik, Bulten, Buonanno, Buskulic, Buy, Byer, Cabero, Cadonati, Cagnoli, Cahillane, Bustillo, Callister, Calloni, Camp, Canepa, Canizares, Cannon, Cao, Cao, Capano, Capocasa, Carbognani, Caride, Carney, Carullo, Diaz, Casentini, Caudill, Cavagli\`a, Cavalier, Cavalieri, Cella, Cepeda, Cerd\'a-Dur\'an, Cerretani, Cesarini, Chaibi, Chamberlin, Chan, Chao, Charlton, Chase, Chassande-Mottin, Chatterjee, Chatziioannou, Cheeseboro, Chen, Chen, Chen, Cheng, Chia, Chincarini, Chiummo, Chmiel, Cho, Cho, Chow, Christensen, Chu, Chua, Chua, Chung, Chung, Ciani, Ciobanu, Ciolfi, Cipriano, Cirelli, Cirone, Clara, Clark, Clearwater, Cleva, Cocchieri, Coccia, Cohadon, Cohen, Colla, Collette, Collins, Cominsky, Constancio, Conti, Cooper, Corban, Corbitt,
  Cordero-Carri\'on, Corley, Cornish, Corsi, Cortese, Costa, Cotesta, Coughlin, Coughlin, Coulon, Countryman, Couvares, Covas, Cowan, Coward, Cowart, Coyne, Coyne, Creighton, Creighton, Cripe, Crowder, Cullen, Cumming, Cunningham, Cuoco, Canton, D\'alya, Danilishin, D'Antonio, Danzmann, Dasgupta, Costa, Dattilo, Dave, Davier, Davis, Daw, Day, DeBra, Deenadayalan, Degallaix, De~Laurentis, Del\'eglise, Del~Pozzo, Demos, Denker, Dent, De~Pietri, Derby, Dergachev, De~Rosa, De~Rossi, DeSalvo, de~Varona, Dhurandhar, D\'{\i}az, Dietrich, Di~Fiore, Di~Giovanni, Di~Girolamo, Di~Lieto, Ding, Di~Pace, Di~Palma, Di~Renzo, Dmitriev, Doctor, Dolique, Donovan, Dooley, Doravari, Dorrington, \'Alvarez, Downes, Drago, Dreissigacker, Driggers, Du, Dudi, Dupej, Dwyer, Easter, Edo, Edwards, Effler, Eggenstein, Ehrens, Eichholz, Eikenberry, Eisenmann, Eisenstein, Essick, Estelles, Estevez, Etienne, Etzel, Evans, Evans, Fafone, Fair, Fairhurst, Fan, Farinon, Farr, Farr, Fauchon-Jones, Favata, Fays, Fee, Fehrmann, Feicht, Fejer,
  Feng, Fernandez-Galiana, Ferrante, Ferreira, Ferrini, Fidecaro, Fiori, Fiorucci, Fishbach, Fisher, Fishner, Fitz-Axen, Flaminio, Fletcher, Fong, Font, Forsyth, Forsyth, Fournier, Frasca, Frasconi, Frei, Freise, Frey, Frey, Fritschel, Frolov, Fulda, Fyffe, Gabbard, Gadre, Gaebel, Gair, Gammaitoni, Ganija, Gaonkar, Garcia, Garc\'{\i}a-Quir\'os, Garufi, Gateley, Gaudio, Gaur, Gayathri, Gemme, Genin, Gennai, George, George, Gergely, Germain, Ghonge, Ghosh, Ghosh, Ghosh, Giacomazzo, Giaime, Giardina, Giazotto, Gill, Giordano, Glover, Goetz, Goetz, Goncharov, Gonz\'alez, Castro, Gopakumar, Gorodetsky, Gossan, Gosselin, Gouaty, Grado, Graef, Granata, Grant, Gras, Gray, Greco, Green, Green, Gretarsson, Groot, Grote, Grunewald, Gruning, Guidi, Gulati, Guo, Gupta, Gupta, Gushwa, Gustafson, Gustafson, Halim, Hall, Hall, Hamilton, Hamilton, Hammond, Haney, Hanke, Hanks, Hanna, Hannam, Hannuksela, Hanson, Hardwick, Harms, Harry, Harry, Hart, Haster, Haughian, Healy, Heidmann, Heintze, Heitmann, Hello, Hemming, Hendry,
  Heng, Hennig, Heptonstall, Hernandez, Heurs, Hild, Hinderer, Hoak, Hochheim, Hofman, Holland, Holt, Holz, Hopkins, Horst, Hough, Houston, Howell, Hreibi, Huerta, Huet, Hughey, Hulko, Husa, Huttner, Huynh-Dinh, Iess, Indik, Ingram, Inta, Intini, Isa, Isac, Isi, Iyer, Izumi, Jacqmin, Jani, Jaranowski, Johnson, Johnson, Jones, Jones, Jonker, Ju, Junker, Kalaghatgi, Kalogera, Kamai, Kandhasamy, Kang, Kanner, Kapadia, Karki, Karvinen, Kasprzack, Kastaun, Katolik, Katsanevas, Katsavounidis, Katzman, Kaufer, Kawabe, Keerthana, K\'ef\'elian, Keitel, Kemball, Kennedy, Key, Khalili, Khamesra, Khan, Khan, Khan, Khan, Khazanov, Kijbunchoo, Kim, Kim, Kim, Kim, Kim, Kim, King, King, Kinley-Hanlon, Kirchhoff, Kissel, Kleybolte, Klimenko, Knowles, Koch, Koehlenbeck, Koley, Kondrashov, Kontos, Korobko, Korth, Kowalska, Kozak, Kr\"amer, Kringel, Krishnan, Kr\'olak, Kuehn, Kumar, Kumar, Kumar, Kuo, Kutynia, Kwang, Lackey, Lai, Landry, Landry, Lang, Lange, Lantz, Lanza, Lartaux-Vollard, Lasky, Laxen, Lazzarini, Lazzaro, Leaci,
  Leavey, Lee, Lee, Lee, Lee, Lee, Lehmann, Lenon, Leonardi, Leroy, Letendre, Levin, Li, Li, Li, Linker, Littenberg, Liu, Liu, Lo, Lockerbie, London, Longo, Lorenzini, Loriette, Lormand, Losurdo, Lough, Lousto, Lovelace, L\"uck, Lumaca, Lundgren, Lynch, Ma, Macas, Macfoy, Machenschalk, MacInnis, Macleod, Hernandez, Maga\~na Sandoval, Zertuche, Magee, Majorana, Maksimovic, Man, Mandic, Mangano, Mansell, Manske, Mantovani, Marchesoni, Marion, M\'arka, M\'arka, Markakis, Markosyan, Markowitz, Maros, Marquina, Martelli, Martellini, Martin, Martin, Martynov, Mason, Massera, Masserot, Massinger, Masso-Reid, Mastrogiovanni, Matas, Matichard, Matone, Mavalvala, Mazumder, McCann, McCarthy, McClelland, McCormick, McCuller, McGuire, McIver, McManus, McRae, McWilliams, Meacher, Meadors, Mehmet, Meidam, Mejuto-Villa, Melatos, Mendell, Mendoza-Gandara, Mercer, Mereni, Merilh, Merzougui, Meshkov, Messenger, Messick, Metzdorff, Meyers, Miao, Michel, Middleton, Mikhailov, Milano, Miller, Miller, Miller, Miller, Millhouse,
  Mills, Milovich-Goff, Minazzoli, Minenkov, Ming, Mishra, Mitra, Mitrofanov, Mitselmakher, Mittleman, Moffa, Mogushi, Mohan, Mohapatra, Montani, Moore, Moraru, Moreno, Morisaki, Mours, Mow-Lowry, Mueller, Muir, Mukherjee, Mukherjee, Mukherjee, Mukund, Mullavey, Munch, Mu\~niz, Muratore, Murray, Nagar, Napier, Nardecchia, Naticchioni, Nayak, Neilson, Nelemans, Nelson, Nery, Neunzert, Nevin, Newport, Ng, Ng, Nguyen, Nguyen, Nichols, Nielsen, Nissanke, Nitz, Nocera, Nolting, North, Nuttall, Obergaulinger, Oberling, O'Brien, O'Dea, Ogin, Oh, Oh, Ohme, Ohta, Okada, Oliver, Oppermann, Oram, O'Reilly, Ormiston, Ortega, O'Shaughnessy, Ossokine, Ottaway, Overmier, Owen, Pace, Pagano, Page, Page, Pai, Pai, Palamos, Palashov, Palomba, Pal-Singh, Pan, Pan, Pang, Pang, Pankow, Pannarale, Pant, Paoletti, Paoli, Papa, Parida, Parker, Pascucci, Pasqualetti, Passaquieti, Passuello, Patil, Patricelli, Pearlstone, Pedersen, Pedraza, Pedurand, Pekowsky, Pele, Penn, Perez, Perreca, Perri, Pfeiffer, Phelps, Phukon, Piccinni,
  Pichot, Piergiovanni, Pierro, Pillant, Pinard, Pinto, Pirello, Pitkin, Poggiani, Popolizio, Porter, Possenti, Post, Powell, Prasad, Pratt, Pratten, Predoi, Prestegard, Principe, Privitera, Prodi, Prokhorov, Puncken, Punturo, Puppo, P\"urrer, Qi, Quetschke, Quintero, Quitzow-James, Raab, Rabeling, Radkins, Raffai, Raja, Rajan, Rajbhandari, Rakhmanov, Ramirez, Ramos-Buades, Rana, Rapagnani, Raymond, Razzano, Read, Regimbau, Rei, Reid, Reitze, Ren, Ricci, Ricker, Riemenschneider, Riles, Rizzo, Robertson, Robie, Robinet, Robson, Rocchi, Rolland, Rollins, Roma, Romano, Romel, Romie, Rosi\ifmmode~\acute{n}\else \'{n}\fi{}ska, Ross, Rowan, R\"udiger, Ruggi, Rutins, Ryan, Sachdev, Sadecki, Sakellariadou, Salconi, Saleem, Salemi, Samajdar, Sammut, Sampson, Sanchez, Sanchez, Sanchis-Gual, Sandberg, Sanders, Sarin, Sassolas, Sathyaprakash, Saulson, Sauter, Savage, Sawadsky, Schale, Scheel, Scheuer, Schmidt, Schnabel, Schofield, Sch\"onbeck, Schreiber, Schuette, Schulte, Schutz, Schwalbe, Scott, Scott, Seidel, Sellers,
  Sengupta, Sentenac, Sequino, Sergeev, Setyawati, Shaddock, Shaffer, Shah, Shahriar, Shaner, Shao, Shapiro, Shawhan, Shen, Shoemaker, Shoemaker, Siellez, Siemens, Sieniawska, Sigg, Silva, Singer, Singh, Singhal, Sintes, Slagmolen, Slaven-Blair, Smith, Smith, Smith, Somala, Son, Sorazu, Sorrentino, Souradeep, Spencer, Srivastava, Staats, Steinke, Steinlechner, Steinlechner, Steinmeyer, Steltner, Stevenson, Stocks, Stone, Stops, Strain, Stratta, Strigin, Strunk, Sturani, Stuver, Summerscales, Sun, Sunil, Suresh, Sutton, Swinkels, Szczepa\ifmmode~\acute{n}\else \'{n}\fi{}czyk, Tacca, Tait, Talbot, Talukder, Tanner, T\'apai, Taracchini, Tasson, Taylor, Taylor, Tewari, Theeg, Thies, Thomas, Thomas, Thomas, Thorne, Thrane, Tiwari, Tiwari, Tokmakov, Toland, Tonelli, Tornasi, Torres-Forn\'e, Torrie, T\"oyr\"a, Travasso, Traylor, Trinastic, Tringali, Trozzo, Tsang, Tse, Tso, Tsuna, Tsukada, Tuyenbayev, Ueno, Ugolini, Urban, Usman, Vahlbruch, Vajente, Valdes, van Bakel, van Beuzekom, van~den Brand, Van Den~Broeck,
  Vander-Hyde, van~der Schaaf, van Heijningen, van Veggel, Vardaro, Varma, Vass, Vas\'uth, Vecchio, Vedovato, Veitch, Veitch, Venkateswara, Venugopalan, Verkindt, Vetrano, Vicer\'e, Viets, Vinciguerra, Vine, Vinet, Vitale, Vo, Vocca, Vorvick, Vyatchanin, Wade, Wade, Wade, Walet, Walker, Wallace, Walsh, Wang, Wang, Wang, Wang, Wang, Ward, Warner, Was, Watchi, Weaver, Wei, Weinert, Weinstein, Weiss, Wellmann, Wen, Wessel, We\ss{}els, Westerweck, Wette, Whelan, Whiting, Whittle, Wilken, Williams, Williams, Williamson, Willis, Willke, Wimmer, Winkler, Wipf, Wittel, Woan, Woehler, Wofford, Wong, Worden, Wright, Wu, Wysocki, Xiao, Yam, Yamamoto, Yancey, Yang, Yap, Yazback, Yu, Yu, Yvert, Zadro\ifmmode~\dot{z}\else \.{z}\fi{}ny, Zanolin, Zelenova, Zendri, Zevin, Zhang, Zhang, Zhang, Zhang, Zhang, Zhao, Zhou, Zhou, Zhu, Zhu, Zimmerman, Zlochower, Zucker, \& Zweizig}]{Abbott_2019}
---. 2019, Physical Review X, 9, 011001, \dodoi{10.1103/physrevx.9.011001}

\bibitem[{Abbott {et~al.}(2020)}]{Abbott_2020}
Abbott, B.~P., {et~al.} 2020, Living Reviews in Relativity, 23, \dodoi{10.1007/s41114-020-00026-9}

\bibitem[{Acernese {et~al.}(2014)}]{VIRGO:2014yos}
Acernese, F., {et~al.} 2014, Classical and Quantum Gravity, 32, 024001, \dodoi{10.1088/0264-9381/32/2/024001}

\bibitem[{Aghanim {et~al.}(2020)}]{Planck:2018vyg}
Aghanim, N., {et~al.} 2020, Astron. Astrophys., 641, A6, \dodoi{10.1051/0004-6361/201833910}

\bibitem[{Alam {et~al.}(2017)}]{BOSS:2016wmc}
Alam, S., {et~al.} 2017, Mon. Not. Roy. Astron. Soc., 470, 2617, \dodoi{10.1093/mnras/stx721}

\bibitem[{Chen {et~al.}(2018)Chen, Fishbach, \& Holz}]{Chen_2018}
Chen, H.-Y., Fishbach, M., \& Holz, D.~E. 2018, Nature, 562, 545–547, \dodoi{10.1038/s41586-018-0606-0}

\bibitem[{Chen {et~al.}(2021)Chen, Holz, Miller, Evans, Vitale, \& Creighton}]{Chen_2021}
Chen, H.-Y., Holz, D.~E., Miller, J., {et~al.} 2021, Classical and Quantum Gravity, 38, 055010, \dodoi{10.1088/1361-6382/abd594}

\bibitem[{{Chen} {et~al.}(2024){Chen}, {Talbot}, \& {Chase}}]{Chen_2020}
{Chen}, H.-Y., {Talbot}, C., \& {Chase}, E.~A. 2024, \prl, 132, 191003, \dodoi{10.1103/PhysRevLett.132.191003}

\bibitem[{Chen {et~al.}(2019)Chen, Vitale, \& Narayan}]{ChenViewingAngle}
Chen, H.-Y., Vitale, S., \& Narayan, R. 2019, Phys. Rev. X, 9, 031028, \dodoi{10.1103/PhysRevX.9.031028}

\bibitem[{DES \& Collaborations(2023)}]{PhysRevD.107.023531}
DES, \& Collaborations, S. 2023, Phys. Rev. D, 107, 023531, \dodoi{10.1103/PhysRevD.107.023531}

\bibitem[{Dhawan {et~al.}(2020)Dhawan, Bulla, Goobar, Sagués~Carracedo, \& Setzer}]{Dhawan_2020}
Dhawan, S., Bulla, M., Goobar, A., Sagués~Carracedo, A., \& Setzer, C.~N. 2020, The Astrophysical Journal, 888, 67, \dodoi{10.3847/1538-4357/ab5799}

\bibitem[{Di~Valentino {et~al.}(2021)Di~Valentino, Mena, Pan, Visinelli, Yang, Melchiorri, Mota, Riess, \& Silk}]{DiValentino:2021izs}
Di~Valentino, E., Mena, O., Pan, S., {et~al.} 2021, Class. Quant. Grav., 38, 153001, \dodoi{10.1088/1361-6382/ac086d}

\bibitem[{Evans {et~al.}(2017)}]{evans}
Evans, P.~A., {et~al.} 2017, Science, 358, 1565, \dodoi{10.1126/science.aap9580}

\bibitem[{{Feeney} {et~al.}(2019){Feeney}, {Peiris}, {Williamson}, {Nissanke}, {Mortlock}, {Alsing}, \& {Scolnic}}]{Feeney_2019}
{Feeney}, S.~M., {Peiris}, H.~V., {Williamson}, A.~R., {et~al.} 2019, \prl, 122, 061105, \dodoi{10.1103/PhysRevLett.122.061105}

\bibitem[{Finstad {et~al.}(2018)Finstad, De, Brown, Berger, \& Biwer}]{GW170817_GWEM_angle}
Finstad, D., De, S., Brown, D.~A., Berger, E., \& Biwer, C.~M. 2018, The Astrophysical Journal Letters, 860, L2, \dodoi{10.3847/2041-8213/aac6c1}

\bibitem[{Foreman-Mackey {et~al.}(2013)Foreman-Mackey, Hogg, Lang, \& Goodman}]{Foreman_Mackey_2013}
Foreman-Mackey, D., Hogg, D.~W., Lang, D., \& Goodman, J. 2013, Publications of the Astronomical Society of the Pacific, 125, 306–312, \dodoi{10.1086/670067}

\bibitem[{Guidorzi {et~al.}(2017)Guidorzi, Margutti, Brout, {et~al.}}]{Guidorzi_2017}
Guidorzi, C., Margutti, R., Brout, D., {et~al.} 2017, The Astrophysical Journal Letters, 851, L36, \dodoi{10.3847/2041-8213/aaa009}

\bibitem[{Heinzel {et~al.}(2021)}]{10.1093/mnras/stab221}
Heinzel, J., {et~al.} 2021, Monthly Notices of the Royal Astronomical Society, 502, 3057, \dodoi{10.1093/mnras/stab221}

\bibitem[{Holz \& Hughes(2005)}]{Holz_2005}
Holz, D.~E., \& Hughes, S.~A. 2005, The Astrophysical Journal, 629, 15–22, \dodoi{10.1086/431341}

\bibitem[{Hotokezaka {et~al.}(2018)Hotokezaka, Nakar, Gottlieb, Nissanke, Masuda, Hallinan, Mooley, \& Deller}]{hotokezaka2018hubble}
Hotokezaka, K., Nakar, E., Gottlieb, O., {et~al.} 2018, Nature Astronomy, 3, \dodoi{10.1038/s41550-019-0820-1}

\bibitem[{Kamionkowski \& Riess(2023)}]{kamionkowski2022hubble}
Kamionkowski, M., \& Riess, A.~G. 2023, Annual Review of Nuclear and Particle Science, 73, 153, \dodoi{https://doi.org/10.1146/annurev-nucl-111422-024107}

\bibitem[{Loredo \& Lamb(2002)}]{Loredo_selection}
Loredo, T.~J., \& Lamb, D.~Q. 2002, Phys. Rev. D, 65, 063002, \dodoi{10.1103/PhysRevD.65.063002}

\bibitem[{Mandel {et~al.}(2019)Mandel, Farr, \& Gair}]{mandel_2019}
Mandel, I., Farr, W.~M., \& Gair, J.~R. 2019, Monthly Notices of the Royal Astronomical Society, 486, 1086, \dodoi{10.1093/mnras/stz896}

\bibitem[{Margutti {et~al.}(2017)Margutti, Berger, Fong, Guidorzi, Alexander, Metzger, Blanchard, Cowperthwaite, Chornock, Eftekhari, Nicholl, Villar, Williams, Annis, Brown, Chen, Doctor, Frieman, Holz, Sako, \& Soares-Santos}]{Margutti2017}
Margutti, R., Berger, E., Fong, W., {et~al.} 2017, The Astrophysical Journal Letters, 848, L20, \dodoi{10.3847/2041-8213/aa9057}

\bibitem[{Mooley {et~al.}(2018)Mooley, Deller, Gottlieb, Nakar, Hallinan, Bourke, Frail, Horesh, Corsi, \& Hotokezaka}]{Mooley_2018}
Mooley, K.~P., Deller, A.~T., Gottlieb, O., {et~al.} 2018, Nature, 561, 355–359, \dodoi{10.1038/s41586-018-0486-3}

\bibitem[{Nissanke {et~al.}(2013)Nissanke, Holz, Dalal, Hughes, Sievers, \& Hirata}]{nissanke2013determining}
Nissanke, S., Holz, D.~E., Dalal, N., {et~al.} 2013.
\newblock \doarXiv{1307.2638}

\bibitem[{Palmese {et~al.}(2024)Palmese, Kaur, Hajela, Margutti, McDowell, \& MacFadyen}]{palmese2023standard}
Palmese, A., Kaur, R., Hajela, A., {et~al.} 2024, Phys. Rev. D, 109, 063508, \dodoi{10.1103/PhysRevD.109.063508}

\bibitem[{Peng {et~al.}(2024)Peng, Risti\ifmmode~\acute{c}\else \'{c}\fi{}, Kedia, O'Shaughnessy, Fontes, Fryer, Korobkin, Mumpower, Villar, \& Wollaeger}]{peng2024kilonova}
Peng, Y., Risti\ifmmode~\acute{c}\else \'{c}\fi{}, M., Kedia, A., {et~al.} 2024, Phys. Rev. Res., 6, 033078, \dodoi{10.1103/PhysRevResearch.6.033078}

\bibitem[{Riess {et~al.}(2016)}]{Riess:2016jrr}
Riess, A.~G., {et~al.} 2016, Astrophys. J., 826, 56, \dodoi{10.3847/0004-637X/826/1/56}

\bibitem[{Riess {et~al.}(2022)Riess, Yuan, Macri, Scolnic, Brout, Casertano, Jones, Murakami, Anand, Breuval, Brink, Filippenko, Hoffmann, Jha, Kenworthy, Mackenty, Stahl, \& Zheng}]{Riess_2022}
Riess, A.~G., Yuan, W., Macri, L.~M., {et~al.} 2022, The Astrophysical Journal Letters, 934, L7, \dodoi{10.3847/2041-8213/ac5c5b}

\bibitem[{Ross {et~al.}(2015)Ross, Samushia, Howlett, Percival, Burden, \& Manera}]{Ross:2014qpa}
Ross, A.~J., Samushia, L., Howlett, C., {et~al.} 2015, Mon. Not. Roy. Astron. Soc., 449, 835, \dodoi{10.1093/mnras/stv154}

\bibitem[{{Schutz}(1986)}]{1986Natur.323..310S}
{Schutz}, B.~F. 1986, \nat, 323, 310, \dodoi{10.1038/323310a0}

\bibitem[{Thrane \& Talbot(2019)}]{Thrane_2019}
Thrane, E., \& Talbot, C. 2019, Publications of the Astronomical Society of Australia, 36, \dodoi{10.1017/pasa.2019.2}

\bibitem[{Verde {et~al.}(2019)Verde, Treu, \& Riess}]{Verde:2019ivm}
Verde, L., Treu, T., \& Riess, A.~G. 2019, Nature Astron., 3, 891, \dodoi{10.1038/s41550-019-0902-0}

\bibitem[{Vitale {et~al.}(2020)Vitale, Gerosa, Farr, \& Taylor}]{Vitale2020}
Vitale, S., Gerosa, D., Farr, W.~M., \& Taylor, S.~R. 2020, 1, \dodoi{10.1007/978-981-15-4702-7_45-1}

\bibitem[{Wang \& Giannios(2021)}]{Wang_2021}
Wang, H., \& Giannios, D. 2021, The Astrophysical Journal, 908, 200, \dodoi{10.3847/1538-4357/abd39c}

\bibitem[{Wu \& MacFadyen(2018)}]{Wu_2018}
Wu, Y., \& MacFadyen, A. 2018, The Astrophysical Journal, 869, 55, \dodoi{10.3847/1538-4357/aae9de}

\end{thebibliography}
\bibliographystyle{aasjournal}

\end{document}